\documentstyle[aps]{revtex}
\input{psfig}
\catcode`@=11
\def\seceqaa{\@addtoreset{equation}{section}
           \def\theequation{A\arabic{equation}}}
\def\seceqbb{\@addtoreset{equation}{section}
           \def\theequation{B\arabic{equation}}}
\def\seceqcc{\@addtoreset{equation}{section}
           \def\theequation{C\arabic{equation}}}

\catcode`@=12
\begin{document}

\title{2-Nucleon 1-Loop Corrections to Pion Double Charge Exchange within 
Heavy Baryon Chiral Perturbation Theory}
\author{{A. Misra$^{(1)}$} \thanks{e-mail: aalok@iitk.ac.in},
{D. S. Koltun$^{(2)}$} \thanks{e-mail: koltun@urhep.pas.rochester.edu}\\
(1) Indian Institute of Technology, Kanpur 208 016, UP, India,\\ 
(2) University of Rochester, Rochester, NY 14627, USA}
\maketitle
\vskip 0.5 true in

\begin{abstract}

One-loop corrections at the two-nucleon level to 
Pion Double Charge exchange (DCX) scattering  off a nuclear target
at threshold,  are calculated within the framework of  
Heavy Baryon Chiral Perturbation Theory (HBChPT). An
estimate for the (2-nucleon) 1-loop correction is obtained in the
static limit and using an impulse approximation. We find
a small (1.6$\%$) increase relative
to the leading order tree graphs.

\end{abstract}

PACS numbers: 12.39.Fe, 13.75.Gx, 25.80.Gn

Keywords: Effective Field Theories, ([Heavy] Baryon) Chiral
Perturbation Theory, Pion-Nucleus Double Charge Exchange

\clearpage
\section{Introduction}

The  effective field theory used most extensively
to study  QCD at low energies is generically referred to as
Chiral Perturbation Theory (ChPT). With the inclusion of
baryons, the effective theory is called Baryon ChPT  (BChPT), whose
nonrelativistic limit (with respect to the baryons) is referred 
to as Heavy BChPT (HBChPT). So far, pion-nucleus scattering and production 
processes involving multiple nucleons with an arbitrary number of pions,  
have been considered within HBChPT up to the tree level, with one-loop 
corrections only at the single-nucleon level (vertex corrections) 
\cite{bn,myhr}. In this paper, we perform a 2-nucleon-1-loop calculation 
involving pion loops, with the pions being emitted and absorbed at different 
nucleons, which we believe has not been done before.

The goal of this paper is to determine the size of the 1-loop contributions 
to pion Double Charge Exchange (DCX) scattering at threshold on a nuclear 
target, relative to the tree graphs, in the framework of HBChPT.  
One of the motivations for this study is the fact 
that sizable 1-loop (pion) contributions to $\pi-\pi$ scattering and 
$\pi$-N scattering have been obtained in the framework of (HB)ChPT by previous 
authors (\cite {gl,bkm3}). So, it's natural to ask whether similar large
contributions are found for a 2-nucleon calculation involving pions in 
HBChPT to one loop. We shall find that this is not the case; the loop
correction to the two nucleon process is small, as expected for a chiral
expansion.

The 2-nucleon process considered is pion DCX:
\begin{equation} 
\label{eq:dcxdef}
\pi^++n+n\rightarrow\pi^-+p+p,
\end{equation}
where the nucleons are in bound nuclear states.
We consider only transitions to the DIAS 
($\equiv $ Double Isobaric Analog States), e.g. 
$^{14}{\rm C}(\pi^+,\pi^-)^{14}{\rm O}$ (DIAS). 
The DIAS  is that (normalized) state  obtained by operating
with $I_+I_+$ on the target ground state, 
where $I$ is the total isospin operator.
The DCX contribution is much smaller than elastic 
scattering ($^{14}$C$\rightarrow^{14}$C) because while the former involves 
only the valence nucleons, the latter involves a coherent scattering of 
the core and the valence nucleons. 

The reason for considering DCX at threshold is that it is dominated
by two-nucleon processes, since single $\pi-N$ scattering cannot contribute.
It has been shown that the contribution of meson exchange currents (MEC)
to DCX, although less than double-scattering, is not all that small.
(See \cite {kj} and references therein.) In the context of HBChPT, these
results are at the leading order, or tree level. The second motivation of
the present paper is to establish the size of the next-order correction to
the two-nucleon MEC contribution to DCX.

In Section II, we introduce the theoretical background of single- 
and multi-nucleon HBChPT and discuss a chiral
power counting rule due to Weinberg (including issues of reducibility
of graphs, which is an additional  complication arising at
the multi-nucleon level). 
In Section III, we discuss the approximations
involved in getting (analytical and numerical) estimates of the amplitudes for 
the tree and 1-loop graphs for pion DCX. The relevant leading
order tree graphs are evaluated in the 
framework of HBChPT and compared to an earlier calculation.  
In Section IV, we  evaluate 
the 2-nucleon 1-loop corrections to the tree graphs of Section III.
We then make numerical estimate of the finite parts of the (2-nucleon) 1-loop
graphs, and make comparison with the tree graphs of Section III. 
In Section V, we discuss the renormalization 
of the 1-loop graphs of Section IV using
2$\pi$ - 2nucleon contact terms. Section VI summarizes and 
discusses the findings for the DCX problem. We include an estimate of the
effect of vertex corrections on the tree graph amplitudes, based on earlier
published work. There are two appendices: Appendix A presents the
vertices written in terms of $\pi^\pm,\pi^0$ [rather than their cartesian
counterparts (as in \cite{bkm1})  as the former are more readily useful
for calculation purposes] (${\bf A.1}$), and combinatoric factors (${\bf A.2}$)
for the 1-loop $\pi$-NN graphs. Appendix B discusses the various 1-loop
integrals  and related identities involving them, relevant to the DCX 1-loop 
calculation, highlighting the ones that are new.

\section{Theoretical Background}

In this section, we discuss the basic elements
of HBChPT at the single and multi-nucleon level that will be required later in 
the paper. 

The leading order (LO) HBChPT Lagrangian that will be  used in
the calculations of the LO tree and and 1-loop graphs is given
by:
\begin{eqnarray}
\label{eq:LOLag}
& & {\bar{\rm H}}(iv\cdot{\rm D}+g_A^0{\rm S}\cdot u){\rm H}
\nonumber\\
& & +{F^2\over4}\biggl(\langle\partial^\mu U^\dagger
\partial_\mu U\rangle
+ M\langle U+U^\dagger-2\rangle\biggr),
\end{eqnarray} 
where $g_A^0\equiv$ axial-vector
coupling constant, $F\equiv$ pion-decay constant and
$M\equiv$pion mass in the chiral limit.  
The trace in the nucleon isospin
space is denoted by $\langle\ \rangle$ in (\ref{eq:LOLag}).
 
The HBChPT Lagrangian is written in terms of the ``upper component"
H and its  covariant adjoint ${\bar{\rm H}}$, 
exponentially parametrized matrix-valued meson fields $U,\ u\equiv\sqrt{U}$, 
baryon (``$v_\mu,\rm S_\nu$") and
pion-field-dependent (``${\rm D}_\mu, u_\nu,\chi_\pm$") building
blocks defined below:
\begin{equation}  
\label{eq:Hdef}
{\rm H}\equiv e^{i{\rm m}v\cdot x}{1\over2}(1+\rlap/v)\psi,
\end{equation}
where $\psi$  is the Dirac spinor
and $m$ is the nucleon mass; 
\begin{eqnarray} 
\label{eq:vSdef}
& & v_\mu\equiv{\rm nucleon\ 4-velocity\ parameter},\nonumber\\
&  & {\rm S}_\nu\equiv{i\over2}\gamma^5\sigma_{\nu\rho}v^\rho
\equiv{\rm Pauli-Lubanski\ spin\ operator};
\end{eqnarray}
\begin{equation}
\label{eq:Udef}
U={\rm exp}\biggl(i{\phi\over F_\pi}\biggr),\ {\rm where}\
\phi\equiv\vec\pi\cdot\vec\tau,
\end{equation}
where $\vec\tau$ are the nucleon isospin generators;
${\rm D}_\mu=\partial_\mu+\Gamma_\mu$ where 
$\Gamma_\mu\equiv {1\over 2}[u^\dagger,\partial_\mu u]$;
$u_{\mu}\equiv i(u^{\dagger}{\partial}_{\mu}u - 
u{\partial}_{\mu}u^{\dagger})$.

The $m\pi - {\bar{\rm N}}$N vertices, $m=1, 3$, 
are constructed from the Yukawa term:
\begin{equation} 
\label{eq:Yukawadef}
g_A^0{\bar{\rm H}}{\rm S}\cdot u^{(1,3)}\rm H,
\end{equation}
(where the superscript on  $u_\mu$ represents the powers of the pion field).
The $2\pi -  {\bar{\rm N}}$N vertex is constructed from the Dirac term:
\begin{equation}
\label{eq:Diracdef}
i{\bar{\rm H}}v\cdot{\rm D}^{(2)}{\rm H}\equiv i{\bar{\rm H}}v
\cdot\Gamma^{(2)}\rm H
\end{equation}
(where the superscripts on $\rm D_\mu$ and 
$\Gamma_\mu$ represent the powers of the pion field). 

The four-pion vertex is constructed from the non-linear sigma model Lagrangian
$\equiv$ LO ChPT Lagrangian:
\begin{equation}
\label{eq:LOChPTdef}
{F^2\over4}\biggl(\langle\partial^\mu U^\dagger\partial_\mu 
U\rangle +M^2\langle U+U^\dagger-2\rangle\biggr).
\end{equation}
For more details, refer to Appendix A.

The elementary vertices from which the tree and the 1-loop
graphs of Sections II and III have been constructed, are drawn in Fig 1.
However, the calculations in the chiral perturbation expansion are
renormalized to the chiral order of the expansion. 
The Weinberg chiral power counting relation (WCPCR, \cite {wnbg1,wnbg2}) 
is used for a systematic classification of the relevant tree and 1-loop graphs.
The relation determines the overall chiral order of irreducible graphs in 
terms of the total number of incoming or outgoing nucleons
(the two  are the same because of baryon number conservation), the total 
number of loops, the chiral order of the vertices,and connectedness of the
graphs, as discussed below. 
Here is WCPCR:
\begin{equation} 
\label{eq:WCPCRdef}
\nu = 4 - N + 2(L - C) +\sum_i v_i({n_i\over 2} + d_i -2),
\end{equation}
where $\nu\equiv$ overall
chiral  order of a graph, $N\equiv$ total number of incoming/outgoing nucleons, 
$L\equiv$ number of loops, $C\equiv$ number of  separately connected pieces of
the graph, $v_i\equiv$ the  number of vertices of type $i$, 
$n_i\equiv$ the number of incoming $and$ outgoing nucleons at the $i$th vertex 
and $d_i\equiv$ is the number of derivatives or powers of $M_\pi$. From 
baryon number conservation, $n_i\equiv2$ or 0. 

Graphs which violate relation (\ref{eq:WCPCRdef}) because of anomalously
small denominators are referred to as reducible by Weinberg \cite {wnbg2}.
This class includes graphs whose energy denominators in old-fashioned 
time-ordered perturbation theory are of the order of ${M_\pi}^2/m$ rather
than $M_\pi$. These graphs arise in the context of $NN$ or many-nucleon
scattering, without external pions. (For a recent treatment of the $NN$
scattering problem which deals with problems of the Weinberg scheme, see
\cite{kap1,kap2}.) However, for pion scattering on $NN$ pairs, as in the 
present paper, all graphs are irreducible, as can be seen from the
discussion of \cite{wnbg2}. In Sections III and IV,
WCPCR will be used to determine the overall chiral order of the tree
and 1-loop graphs.

The lowest order tree graphs for DCX, based on the LO vertices 
(\ref{eq:Yukawadef})-(\ref{eq:LOChPTdef}), will 
be shown to be of chiral order $\nu = 0$ (see Sec. III.B). The one-loop 
corrections at the $2N$ level, which are the main subject of this paper, are 
of order $\nu = 2$ (see Sec. IV). To this same order, there are also 
corrections to the LO vertices, which have already been obtained in (HB)ChPT
(see \cite {gl,bkm3,bkm2})
as renormalized effective interactions, with a number of low energy constants
(LECs). These LECs are to be fixed from experiment, but are not in fact all
known. (For the $\pi-\pi$ vertex, the LECs of $\nu = 2$ are `almost' all
determined \cite{gl}. For $\pi-N$, the information is less complete, but
has been supplemented by theoretical arguments \cite{bkm3,bkm1,bkm2} for
on-shell nucleons.)

However, we do know the renormalized values 
of the pion and nucleon masses, the axial-vector coupling constant, and the
pion-decay constant. For the purpose of numerical calculation of 
the analytical expressions for the tree and 1-loop 
graphs later in the paper, we use
\begin{equation} 
\label{eq:MFgnum}
M_\pi=139.57\ {\rm MeV}, \ F_\pi=93\ {\rm MeV},\ g_A=1.26.
\end{equation}

Renormalizing these constants, including the nucleon mass $m$, to 1-loop
(in the $\pi-N$ and $\pi-\pi$ interactions) gives the following relations
to the corresponding
quantities in the chiral limit $g_A^0, m^0,  M, F$ : 
\begin{eqnarray}
\label{eq:eq1}
& & g_A=g_A^0[1+\rho_g M^2]\nonumber\\
& & m=m^0[1+\rho_m M^2]\nonumber\\
& & M_\pi^2=M^2[1+\rho_M M^2]\nonumber\\
& & F_\pi=F[1+\rho_F M^2],
\end{eqnarray}
where the $\rho_i$ include some of the (HB)ChPT LECs.

It is also known that the 1-loop corrections to the $\pi-\pi$ vertex give
contributions of about 25$\%$ of order $\nu = 2$. Similar corrections to
the $m\pi-N$ vertices (with $m=1,2,3$) are as large as 10 - 30$\%$ \cite
{gl,bkm3,bkm2}. So we know the order of magnitude of the $\nu = 2$
corrections to the LO DCX (tree) amplitudes from the 1-loop vertex
contributions, but cannot determine these corrections completely, without
the undetermined LECs.

For that reason, we shall use the vertices 
(\ref{eq:Yukawadef}) - (\ref{eq:LOChPTdef}), with the renormalized
values (\ref{eq:MFgnum}) for the tree calculation in Sec. III, 
omitting unknown (but 
significant) corrections of the same order. The 1-loop corrections at the 
$2N$ level are calculated in Sec IV with the same vertices.

\section{DCX Scattering Amplitudes; Tree Graphs}

In this section we discuss the approximations that will
be made in evaluation of tree and 1-loop graphs for
DCX scattering of pions off a nuclear target, 
set up the notations and calculate
the amplitudes for the leading order
(DCX) tree graphs.
 
\subsection{Notations and Approximations}

The scattering matrix element $S_{fi}$ is defined as:
\begin{equation} 
\label{eq:Sfidef}
S_{fi}=-i(2\pi)^4\delta^{(4)}(P_f-P_i)
{1\over{(2\pi)^9M_\pi}}{\cal M}.
\end{equation}

In (\ref{eq:Sfidef}), the nuclear scattering amplitude
${\cal M}$ is defined as the matrix element of
a two-body operator $T$: 
\begin{eqnarray}
\label{eq:Tdef}
& &  {\cal M}\equiv\langle\psi_o,\ I=1,\ I_3=1|T
\tau_+^{(1)}\tau_+^{(2)}|\psi_i,\ I=1,\ I_3=-1\rangle\nonumber\\
& & =\langle\psi_o|T|\psi_i\rangle\times
\langle I=1, I_3=1|\tau_+^{(1)}\tau_+^{(2)}|I=1, I_3=-1\rangle,
\end{eqnarray}
where we assume  a target with  $2n +$ isoscalar core ($I=1, I_3=-1)$;
then the DIAS has  $2p +$ isoscalar core ($I=1, I_3=1)$.
The isospin matrix element in the second line of (\ref{eq:Tdef}) 
equals unity, and
will be omitted in the following.

If $p_{1,2}^\mu$ were the 4-momenta of the 
incoming nucleons, $p_{3,4}^\mu$ the 4-momenta of the outgoing nucleons, and 
$q_{1,2}^\mu$ the 4-momenta of the incoming and outgoing pions (respectively)
then
\begin{equation}
\label{eq:M}
 {\cal M}=\int \prod_{i=1}^4 d^3p_i
\langle\psi_o(p_3,p_4)|T|\psi_i(p_1,p_2)\rangle.
\end{equation} 
The nucleons are in a relative $l=0$ state ($l=0$
state being the dominant partial wave for ground states)
and hence from Pauli's exclusion principle, 
the spin of the incoming and outgoing nuclear states must be 
equal to zero.  Hence, from here on, for both
tree and 1-loop graphs, we will simplify the structure of the 
transition operator T assuming that eventually one is going to take it's 
expectation value with respect to $|l=S=0\rangle$ nuclear 
states. 

The following will be used extensively in the same context.
\begin{equation}
\label{eq:S=0me}
 \langle S=0|(\vec\sigma^{(1)}\cdot\vec q)
(\vec\sigma^{(2)}\cdot\vec q)|S=0\rangle
=\langle\vec\sigma^{(1)}\cdot\vec\sigma^{(2)}\rangle_{S=0}
{{\vec q}^2\over3}=-{\vec q}^2.
\end{equation}

Following \cite{bn} and \cite{myhr}: First, the velocity parameters
of the two participating nucleons are both chosen to have the static limit
values, with only a non-vanishing time component , i.e. 
$v^\mu_1=v^\mu_2=(1,\vec 0)$. Second, the nucleons will be 
treated as if they were on-shell  (sometimes referred to as impulse 
approximation). In the HBChPT  formalism,
$p^0$ denotes only the contribution to the time component of the
total nucleon momenta ($\equiv mv+p$),
$\it in\ addition$ to the rest mass energy $m$ (for  the choice of the
nucleon velocity to possess only a non-zero time component). 
In the present  case $p^0=E_B$, which as stated above, we drop.  
Then if we go to the c.m. frame of the nucleons:
\begin{equation} 
\label{eq:cm}
p_1^\mu=(0,\vec p);\ p_2^\mu=(0,-\vec p);\ 
p_3^\mu=(0,\vec p^\prime);\ p_4^\mu=(0,-\vec p^\prime);\ 
q_1^\mu=q_2^\mu= (M_\pi,\vec 0).
\end{equation}
(Note: The external pions are at zero kinetic energy [threshold].) 

For this paper, we shall only  evaluate 
$\langle S=0|T|S=0\rangle$. ${\cal M}$ will be estimated by calculating
$\langle S=0|T|S=0\rangle$ at a typical point:
\begin{equation}
\label{eq:kinptdef}
\vec P^2=M_\pi^2,
\end{equation}
where $\vec P\equiv \vec p^\prime-\vec p$.
This is a reasonable kinematic  point because the inter-nucleon separation
in a nucleus averaged over the nuclear wavefunction is roughly 
$M_\pi^{-1}$. Then 
\begin{eqnarray} 
\label{eq:Mkinpt}
& & 
{\cal M} \approx\langle S=0|T(\vec P^2=M_\pi^2)|S=0\rangle
\int \prod_{i=1}^4 d^3p_i\langle\psi_o(p_3;\ p_4)|\psi_i(p_1;\ p_2)\rangle
\nonumber\\
& & =\langle S=0|T(\vec P^2=M_\pi^2)|S=0\rangle
\end{eqnarray}
where the last line follows because the
overlap integral is unity for $\psi_o\equiv$ DIAS of $\psi_i$.

\subsection{Tree Graphs}

The connected tree graphs are of overall O$(q^0)$, as follows
from the chiral counting law (\ref{eq:WCPCRdef}),
because for them $N=2, L=0, C=1, n_i=2$ or 0, $d_i=1$ or 2. 
The tree graphs have been constructed from the first 
three and the fifth elementary vertices of Fig 1. They are
drawn in Fig 2. For LO graphs, all the vertices have to be of LO, i.e. 
O$(q)\ m\pi-{\bar{\rm N}}{\rm N}\ (m=1,2,3)$ and O$(q^2)$ 4 $\pi$ vertices. 
Using the exponential parameterization for the matrix-valued meson field,
${\cal L}_{\rm HBChPT}$ and ${\cal L}_{\rm ChPT}$ are written down 
explicitly in terms of the pion triplet fields : $\pi^0, \pi^\pm$ and 
p, n fields [refer to Appendix ${\bf A.1}$].

The S-matrix for (\ref{eq:dcxdef}) can be calculated in perturbation theory
using standard Feynman-diagram techniques. 
For the nth order term , $S^{(n)}$, a combinatoric factor $f$ is defined via:
\begin{eqnarray}
\label{eq:Sndef}
 & & S^{(n)} = {i^n\over{n!}}\int\prod_{k=1}^n d^4x_k
{\cal T}
\biggl(\prod_{i=1}^n\sum_j {\cal L}^I_j(x_i)\biggr)\nonumber\\
& & 
=fi^n\int\prod_{k=1}^n d^4x_k {\cal T}\biggl(\prod_{i=1}^n{\cal L}^I_i(x_i)
\biggr)\ ,
\end{eqnarray} 
(${\cal T}\equiv$ time-ordering  operator),
where one uses: ${\cal T}(AB)={\cal T}(BA)$, where $A,B\equiv$ 
bosonic fields or fermionic bilinears.

One can show  that the combinatoric
factors for tree graphs of Fig 2 are:
\begin{eqnarray} 
\label{eq:treefigs}
&  & {\rm contact\ graph\ Fig\ 2a:\  f=1};\nonumber\\ 
& & {\rm Pole\ graph\ Fig\ 2b\:\ f={1\over 2}};\nonumber\\
&  & {\rm double-scattering\ graph\ Fig\ 2c: f={1\over 2}}.
\end{eqnarray}

The amplitudes for the tree graphs are written in terms of Pauli spinors
(to which  H and ${\bar{\rm H}}$ reduce in the static limit) and
the pionic field $\phi^-\tau_+$ (See (\ref{eq:phi1})). 

Using also (\ref{eq:cm}), (\ref{eq:kinptdef}) and (\ref{eq:treefigs}),
the amplitudes for the tree graphs are expressed as operators written
in terms of ${\bf 1}$ and
$\vec\sigma^{(1)}\cdot\vec P\vec\sigma^{(2)}\cdot\vec P$.

What follows are expressions for 
$T$ and $\langle S=0|T(\vec P^2=M_\pi^2)|S=0\rangle$.

(a) Contact graph Fig 2a:

\begin{equation}
\label{eq:Ta}
 T_{[a]} = 2(\sqrt{2})^4{{g_A^2}\over{24F_\pi^4}}
{{\vec\sigma^{(1)}\cdot\vec P\vec\sigma^{(2)}\cdot\vec P}\over{M_\pi^2
+\vec P^2}},
\end{equation}
which using (\ref{eq:S=0me}) and (\ref{eq:kinptdef}), gives:
\begin{equation}\langle S=0|T_{[a]}(\vec P^2={M_\pi^2})|S=0\rangle 
= -{{g_A^2}\over{6F_\pi^4}}
= -1.42\times 10^{-8}\rm MeV^{-4}.
\end{equation}

(b) Pole graph Fig 2b: 

\begin{equation}
\label{eq:Tb}
 T_{[b]} = (\sqrt{2})^6
{{g_A^2}\over{24F_\pi^4}}[2M_\pi^2-\vec P^2]
{{\vec\sigma^{(1)}\cdot\vec P\vec\sigma^{(2)}
\cdot\vec P}\over{[M_\pi^2+\vec P^2]^2}},
\end{equation}
which using (\ref{eq:S=0me}) and (\ref{eq:kinptdef}), gives:
\begin{equation} \langle S|T_{[b]}(\vec P^2=M_\pi^2)|S=0\rangle
= -{{g_A^2}\over{12F_\pi^4}}
= -6.3\times10^{-9}\rm MeV^{-4}.
\end{equation}

(c) Double-scattering graph Fig 2c: 

\begin{equation}
\label{eq:Tc}
T_{[c]} = -(\sqrt{2})^4{M_\pi^2\over{8F_\pi^4}}
{1\over{\vec P^2}},
\end{equation}
which using (\ref{eq:kinptdef}), gives:
\begin{equation} \langle S=0|T_{[c]}(\vec P^2=M_\pi^2)|S=0\rangle
= -{1\over{2F_\pi^4}}
= -2.67\times10^{-8}\rm MeV^{-4}.
\end{equation}

Thus, the total tree-graph amplitude is :
\begin{equation}
\label{eq:Ttotal}
 \langle S=0|T_{total}[tree](\vec P^2=M_\pi^2)|S=0\rangle
 =
-{1\over {4F_\pi^4}}\biggl(g_A^2+2\biggr)
= -4.81\times10^{-8}\rm MeV^{-4}.
\end{equation}

An older form of chiral Lagrangian that predates QCD was given by Olsson and 
Turner (OT) (\cite {ot}), and has been used for tree calculations, e.g., of
DCX \cite{kj}. It consists of the minimum number of derivatives
of the pion and nucleon fields, with two undetermined (model dependent) 
parameters. It can be easily shown that the OT effective Lagrangian is
equivalent to the LO vertices obtained from (\ref{eq:Yukawadef}) 
and (\ref{eq:LOChPTdef}),
with the OT parameters taken to be
``$\xi,\eta$"=$({2\over3},-{1\over 6})$.
Therefore, the tree graph calculation of this section should agree (as
they indeed do) with the expressions obtained in \cite {kj} for the transition 
operators for "forward scattering" for the ``pion-contact" and ``pion-pole" 
graphs, using the same values of the OT parameters. This provides a check on 
the tree-level calculation using HBChPT.

\section{1-Loop Graphs for $\pi$-NN}

In this section, we  discuss how to evaluate
one-loop corrections to the 
tree graphs (that were evaluated in Section III), for pion DCX,
in the framework of  HBChPT.
Loop graphs, in the present context, involve  
the emission of two or more pions from one nucleon leg and their
absorption at another nucleon leg (for multi-nucleon processes). 
For evaluation of the 1-loop integrals that occur in the 2-nucleon-1-loop 
graphs, use has been made of dimensional regularization in which the space-time 
dimension ``$d$" is allowed to vary continuously, and expressions 
obtained after integration, are expanded around $d=4$. 

One gets 
eight 1-loop graphs using all five of the elementary 
$m\pi-{\bar{\rm N}}{\rm N}\ (m=1, 2, 3, 4)$ and 4-$\pi$ vertices of Fig 1. 
They are drawn in
Figs 3 and 4. The uncrossed counterpart of Fig 3d, is not included because it 
belongs to the class of multiple-scattering graphs involving at least 
three nucleons, which are not considered in this paper. For Fig 4h, if 
the two pions exchanged were $\pi^0$'s, then
the amplitudes  would vanish in the static limit. 

The leading order connected 2-nucleon 1-loop graphs (for DCX)  are of  
O$(q^2)$, as can be seen from
(\ref{eq:WCPCRdef}), because for these graphs,
$N=2,\ L=C=1,\ (n_i,\ d_i)=(2,\ 1)\ \rm or\ (0,\ 2)$.
The vertices corresponding to the interaction
of an  even number of pions (two or four for our purpose) with nucleons, 
are obtained from the Dirac term (\ref{eq:Diracdef}), while the vertices 
corresponding to the interaction of an odd number of pions (one or three 
for our purpose) with 
nucleons, are obtained from the Yukawa term (\ref{eq:Yukawadef}).
The LO($\equiv$O$(q^2)$) ChPT Lagrangian 
(\ref{eq:LOChPTdef}), which is equivalent to the non-linear 
$\sigma$ model, is used for  the 4$\pi$ vertex.
However, we do use the renormalized constants 
(\ref{eq:MFgnum}), which introduces some
corrections of higher order. 
For details refer to Appendix ${\bf A.1}$.

The one-loop amplitudes are written using standard Feynman rules. 
The expressions  for the T-matrix elements for the eight one-loop
graphs for fixed momenta of the external nucleons legs, are given below. 
For notational convenience, 
${\rm H}_{\rm neutron}(p_{1,2})$ is represented
as ${\rm n}(p_{1,2})$, and ${\bar{\rm H}}_{\rm proton}(p_{3,4})$ is 
represented as ${\bar{\rm p}}(p_{3,4})$. Including
the combinatoric factors 
``$f$" of (\ref{eq:Sndef}), one arrives at the expressions below 
for $T_{[j]},\ j=a,...,h$. They are first written in a covariant  notation
except that the two velocities of the two nucleons are chosen to be the
same, i.e. $v_1=v_2\equiv v$ (as in Section III). Then the static limit 
kinematics (\ref{eq:cm}) is applied.
(The combinatoric factors are given first for each diagram, and are obtained 
in Appendix ${\bf A.2}$).

(a)
\begin{eqnarray}
\label{eq:Taloop}
T_{[a]} & = & 
{({\sqrt{2})^6g_A^2}\over{96F_\pi^6}}{1\over i}\int
{{d^dk}\over{(2\pi)^d}}\nonumber\\
& & \times\Biggl[{{{\bar{\rm p}}(p_3)v\cdot(2p_2-2p_4){\rm n}(p_1)
{\bar{\rm p}}(p_4){\rm S}^{(2)}\cdot(p_4-p_2+k){\rm S}^{(2)}\cdot k\rm n(p_2)}
\over{(v\cdot(k-p_2)-i\epsilon)(M_\pi^2-k^2-i\epsilon)(M_\pi^2-
(p_4-p_2+k)^2-i\epsilon)}}\Biggr]\ .\nonumber\\
& & 
\end{eqnarray}

Using
(\ref{eq:cm}), (\ref{eq:Taloop}) 
vanishes because $v\cdot(p_2-p_4)\rightarrow$ 0. 

(b)
\begin{eqnarray} 
\label{eq:Tbloop}
& & T_{[b]} = 
2{{(\sqrt{2})^6g_A^2}\over{96F_\pi^6}}{1\over i}\int 
{{d^dk}\over{(2\pi)^d}}\nonumber\\
& & \times\Biggl[{{{\bar{\rm p}}(p_3){\rm S}^{(1)}\cdot k{\rm S}^{(1)}
\cdot(q_1-q_2+2p_4-2p_2+2k)n(p_1)
{\bar{\rm p}}(p_4)v\cdot(p_4-p_2+2k)n(p_2)}\over{(v\cdot(k-p_3)-i\epsilon)
(M_\pi^2-k^2-i\epsilon)[M_\pi^2-(k+p_4-p_2)^2-i\epsilon]}}
\nonumber\\
& & + {{{\bar{\rm p}}(p_3){\rm S}^{(1)}\cdot(q_1-q_2+2p_4-2p_2-2k)
{\rm S}^{(1)}\cdot kn(p_1)
{\bar{\rm p}}(p_4)v\cdot(p_4-p_2-2k)n(p_2)}\over{(v\cdot(k-p_1)-i\epsilon)
(M_\pi^2-k^2-i\epsilon)[M_\pi^2-(k+p_2-p_4)^2-i\epsilon]}}
\Biggr]\ .\nonumber\\
& & 
\end{eqnarray}

(c)
\begin{eqnarray} 
\label{eq:Tcloop}
& & T_{[c]} = 
{{(\sqrt{2})^8g_A^2M_\pi^2}
\over{96F_\pi^6}}{1\over i}\int {{d^dk}\over{(2\pi)^d}}\nonumber\\
& & \times\Biggl[
{{{\bar{\rm p}}(p_3) {\rm S}^{(1)}\cdot(p_3-p_1+k){\rm S}^{(1)}\cdot k
n(p_1)}\over{(v\cdot (k-p_1) - i\epsilon)}}
+{{{\bar{\rm p}}(p_3){\rm S}^{(1)}\cdot k{\rm S}^{(1)}\cdot(p_3-p_1+k)
n(p_1)}
\over{(v\cdot(k+p_3)-i\epsilon)}}\Biggr]\nonumber\\
& & \times\Biggl[{{[4q_1\cdot q_2+2k\cdot(q_1-q_2+k)]
{\bar{\rm p}}(p_4)v\cdot(q_1-q_2+p_3-p_1+2k)\rm n(p_2)}
\over{[M_\pi^2-(k+q_1-q_2)-i\epsilon](M_\pi^2-k^2-i\epsilon)
[M_\pi^2-(k-[p_1-p_3])^2
-i\epsilon]}}\Biggr].
\end{eqnarray}

(d)
One can show that the contribution
of (d)(1) and (d)(2) (in Fig 3d) are equal, giving
an overall  factor of 2 :

\begin{eqnarray} 
\label{eq:Tdloop}
 T_{[d]} & = & 
-{2\over 3}{{g_A^2(\sqrt{2})^6}\over{256F_\pi^6}}
{1\over i}\int {{d^dk}\over{(2\pi)^d}}\nonumber\\
& & \times\Biggl[{{{\bar{\rm p}}(p_3)v\cdot(q_1-p_2+p_4-k)
v\cdot(k-q_2)n(p_1)
{\bar{\rm p}}(p_4)v\cdot(p_2-p_4+2k)\rm n(p_2)}\over{(v\cdot(k+q_2-p_1)-
i\epsilon)(M_\pi^2-k^2-i\epsilon)[M_\pi^2-(k-[p_4-p_2])^2
-i\epsilon]}}\Biggr]\ .\nonumber\\
& & 
\end{eqnarray}

(e)
\begin{eqnarray} 
\label{eq:Te}
 T_{[e]} & = 
& {{(\sqrt{2})^8}\over{384F_\pi^6}}{1\over i}\int 
{{d^dk}\over{(2\pi)^d}}\nonumber\\
& & \times\biggl[{{[4q_1\cdot q_2+2k\cdot(q_1-q_2+k)]}
\over{(M_\pi^2-k^2-i\epsilon)
[M_\pi^2-(q_1-q_2+k)^2-i\epsilon][M_\pi^2-(k-[p_1-p_3])^2
-i\epsilon]}}\biggr]\nonumber\\
& & \times\Biggl[{\bar{\rm p}}(p_3)v\cdot
(p_1-p_3-2k){\rm n}(p_1){\bar{\rm p}}(p_4)
v\cdot(q_1-q_2+p_3-p_1+2k){\rm n}(p_2)\Biggr]\ .\nonumber\\
& & 
\end{eqnarray}

(f)
\begin{eqnarray}  
\label{eq:Tfloop}
 T_{[f]} & = & 
2{{(\sqrt{2})^6}\over{384F_\pi^6}}{1\over i}\int {{d^dk}\over{
(2\pi)^d}}\nonumber\\
& & \times
\biggl[{{{\bar{\rm p}}v\cdot(q_2-q_1+2p_2-2p_4+2k){\rm n}(p_1)
{\bar{\rm p}}(p_4)v\cdot(p_2-p_4+2k){\rm n}(p_2)}\over{(M_\pi^2-k^2
-i\epsilon)[M_\pi^2-(k-[p_4-p_2])^2-i\epsilon]}}\biggr] .\nonumber\\
& & 
\end{eqnarray}

(g)

\begin{eqnarray} 
\label{eq:Tgloop}
& &  T_{[g]} \nonumber\\
& & = -{{g_A^2(\sqrt{2})^6}\over{144F^6}}{1\over i}\int
{{d^dk}\over{(2\pi^d}}\Biggl({{{\bar{\rm p}}(p_3)
{\rm S}^{(1)}\cdot(p_1-p_3+q_1){\rm n}(p_1){\bar{\rm p}}(p_4){\rm S}^{(2)}
\cdot(p_1-p_3+q_1){\rm n(p_2)}}
\over{(M_\pi^2-k^2-i\epsilon)[M_\pi^2
-(k-[q_1+p_1-p_3])^2-i\epsilon]}}\nonumber\\
& & +{{{\bar{\rm p}}(p_3){\rm S}^{(1)}\cdot(p_3-p_1+3k){\rm n}(p_1)
{\bar{\rm p}}(p_4){\rm S}^{(2)}\cdot(2p_2-2p_4-3q_2+3k){\rm n}(p_2)}
\over{(M_\pi^2-k^2-i\epsilon)[M_\pi^2-(k-[q_1+p_1-p_3])^2-
i\epsilon]}}\Biggr)\ .
\end{eqnarray}

(h)
\begin{eqnarray} 
\label{eq:Thloop}
& &  T_{[h]} \nonumber\\
& & = 2{{g_A^2(\sqrt{2})^6}\over{96F_\pi^6}}
{1\over i}\int {{d^dk}\over{(2\pi)^d}}\nonumber\\
& & \times\Biggl({{{\bar{\rm p}}(p_3){\rm S}^{(1)}\cdot(p_2-p_4+q_1-q_2
+3k){\rm n}(p_1){\bar{\rm p}}(p_4){\rm S}^{(2)}\cdot k v\cdot
(k+p_2-p_4-2q_2){\rm n}(p_2)}\over{(v\cdot(k-p_4)-i\epsilon)
(M_\pi^2-(k-[p_4-p_2+q_2])^2-i\epsilon](M_\pi^2-k^2
-i\epsilon)}}\nonumber\\
& & -{{{\bar{\rm p}}(p_3){\rm S}^{(1)}\cdot(p_2-p_4+q_1-q_2-3k){\rm n}(p_1)
{\bar{\rm p}}(p_4)
v\cdot(p_4-p_2-2q_1+k)
{\rm S}\cdot k{\rm n}(p_2)}\over{(v\cdot(k-p_2)-i\epsilon)
(M_\pi^2-k^2-i\epsilon)[M_\pi^2-(k-[q_1+p_2-p_4])^2-i\epsilon]}}
\Biggr)\nonumber\\
& & +q_1\leftrightarrow -q_2\nonumber\\
\end{eqnarray}

The amplitudes for the eight 1-loop graphs are now written in 
terms of the 11 of the 13 1-loop integrals discussed in 
equations $(B1) - (B3)$ of Appendix B. 
They, like the tree graphs of Section III, are evaluated at (\ref{eq:kinptdef}). 
The Pauli matrices $\vec\sigma^{(1),(2)}$ 
come from the non-relativistic reduction of ${\rm S}^{(1),(2)}_\mu$.
The expressions
below are  written in terms of Pauli spin operators, and will be
evaluated for $\langle S=0|T|S=0\rangle$. 
For graphs (a) - (f), the spin operator is unity; for (g) and (h),
we use (\ref{eq:S=0me}).

Using dimensional regularization, the ``$L$" in the 
expressions below is the UV-divergent portions of  the loop integrals and 
is defined by :
\begin{equation} 
\label{eq:Ldef}
L \equiv {\mu^{d-4}\over{16\pi^2}}\biggl({1\over{d-4}}
-{1\over 2}\bigg[1+\Gamma^\prime(1) + ln(4\pi)\biggr]\biggr),
\end{equation}
where $d$ is  the dimension parameter and $\mu$ is the renormalization point,  
which is taken to be of the order
of the nucleon mass. Following \cite{bkm2} we use 
$\mu=1\ {\rm GeV}$, and also vary the value by $\pm0.5$ GeV to test the
sensitivity of the results to this choice.

The momentum loop integrals are
then rewritten in terms of the 1-loop integrals of (B1) - (B3): the 2$\pi$ -
propagator integrals are represented
by ${\cal J}^{\pi\pi}, 
{\cal J}^{\pi\pi}_{1,2,3}$ and the 1-nucleon - 2$\pi$ - propagator integrals 
are represented
by $\gamma_{0,1,2,3,4,5}$.
\footnote{The notations of \cite{bkm1} and \cite{gss}, where similar integrals 
have been evaluated, have been modified for this
paper; see Appendix B.}
 For convenience, they are represented by the
14 integrals $I_i,\ i=1\dots 14$ 
in the one-loop amplitudes (\ref{eq:Taloop1}) - (\ref{eq:Thloop1}).
Then using the results of Appendix ${\bf B.1} - {\bf B.2}$, 
the integrals  and hence the 
amplitudes are written down  as a linear combination of the UV-finite 
(L-independent) and UV-divergent (L-dependent) terms. 
In Table I we identify the equation in Appendix B used to evaluate
each of the 14 1-loop integrals. The numerical values of the UV-finite
parts are also given in Table I, for completeness, so that the calculation 
can be reconstructed by the reader, or
for application of the relevant parts to problems other than DCX.

\begin{table}[htbp]
\centering
\caption{List of 1-Loop Integrals in (\protect\ref{eq:Taloop1}) - 
(\protect\ref{eq:Thloop1})}
\begin{tabular} {|c|c|c|c|} \hline
$I_k$ & Integrals of App B 
& Equation & Numerical value \\ 
& & & of UV-finite  part \\ 
& & & at $\mu=$1 GeV \\ \hline
$I_1$ & 
${\cal J}^{\pi\pi}(-M_\pi^2,M_\pi^2)$ & (B9) & 0.018 \\ \hline 
$I_2$ &  ${\cal J}^{\pi\pi}_1(-M_\pi^2, M_\pi^2)$ & (B19) & 0.009
\\ \hline
$I_3$ &  ${\cal J}^{\pi\pi}_2(-M_\pi^2, M_\pi^2)$ & (B23) 
& 0.014$M_\pi^2$
\\ \hline
$I_4$ &  ${\cal J}^{\pi\pi}_3(-M_\pi^2, M_\pi^2)$ & (B25) & 0.006 
\\ \hline
$I_5$ & ${\partial\over{\partial M^2}}{\cal J}^{\pi\pi}_1
(-M_\pi^2,M^2)_{M^2=M_\pi^2}$ 
& (B21) & $-0.004M_\pi^{-2}$ \\ \hline
$I_6$ & ${\partial\over{\partial M^2}}{\cal J}^{\pi\pi}_2
(-M_\pi^2, M^2)|_{M^2=M_\pi^2}$ & (B27) & 0.002 \\ \hline
$I_7$ & ${\partial\over{\partial M^2}}{\cal J}^{\pi\pi}_3
(-M_\pi^2, M^2)|_{M^2=M_\pi^2}$ & (B29) & -0.004 \\ \hline
$I_8$ & ${\cal J}^{\pi\pi}(0,M_\pi^2)$
& (B10) & 0.019 \\ \hline
$I_9$ & 
${\cal J}^{\pi\pi}_2(0, M_\pi^2)$ & (B8,B17) & 0.013\\ \hline
$I_{10}$ & ${\cal J}^{\pi\pi}_3(0, M_\pi^2)$ &  (B10,B18) & 0.006 \\ \hline
$I_{11}$ & $\gamma_0(-M_\pi, 0, -M_\pi^2)$ & (B13) 
& $0.019M_\pi^{-1}$ \\ \hline 
$I_{12}$ & $\gamma_2(0,-M_\pi,0) -\gamma_2(0,M_\pi,0)$ & (B31)
& $0.025M_\pi^{-1}$ \\ \hline
$I_{13}$ & $\gamma_3(0,-M_\pi,0)-\gamma_3(0,M_\pi,0)$ 
& (B34) & -0.022$M_\pi$ \\ \hline
$I_{14}$ & $\gamma_5(0,-M_\pi,0)-\gamma_5(0,M_\pi,0)$ & (B37) 
& $0.022M_\pi^{-1}$
 \\ \hline
\end{tabular}
\end{table}

\clearpage

(a)
\begin{equation}
\label{eq:Taloop1}
 T_{[a]}=0
\end{equation}
because $v\cdot(p_2-p_4)=0$. 

(b)
\begin{eqnarray} 
\label{eq:Tbloop1}
 T_{[b]} & = & 2(\sqrt{2})^6{{g_A^2}
\over{48F^6}}\biggl[-3I_3+
M_\pi^2I_4 -M_\pi^2I_2\biggr]\nonumber\\
& & = 2(\sqrt{2})^6{{g_A^2M_\pi^2}\over{48F_\pi^6}}\biggl[
{{29}\over{6}}L-{85\over{576\pi^2}}+{{\sqrt{5}}\over{8\pi^2}}
ln\biggl({{\sqrt{5}+1}\over{\sqrt{5}-1}}\biggr)\nonumber\\
& & +{{29}\over{192\pi^2}}ln{M_\pi^2\over\mu^2}\biggr]
\end{eqnarray}

(c)
\begin{eqnarray}
\label{eq:Tcloop1}
 T_{[c]} & = & 
{{g_A^2(\sqrt{2})^8}\over{96F_\pi^6}}
\Biggl[-6 M_\pi^2
\biggl( -M_\pi^2I_5-3I_6 +M_\pi^2I_7\biggr)\nonumber\\
& & +2\biggl(M_\pi^2I_2+3I_3 -M_\pi^2I_4\biggr)\Biggr]\nonumber\\
& & = {{g_A^2M_\pi^2(\sqrt{2})^8}\over{96F_\pi^6}}\nonumber\\
& & \times\Biggl(-{{41}\over 3}L
-{{23}\over{288\pi^2}}-{{41\sqrt{5}}\over{120\pi^2}}
ln\biggl({{\sqrt{5}+1}\over{\sqrt{5}-1}}\biggr)
-{25\over{48\pi^2}}ln{M_\pi^2\over\mu^2}\biggr)
\end{eqnarray}

(d)
\begin{eqnarray} 
\label{eq:Tdloop1}
T_{[d]} & = & 
-2{{(\sqrt{2})^6}\over3}{1\over{128F_\pi^6}} 
\biggl[-I_3 -4M_\pi^2I_1 + 4M_\pi^3I_{11}\biggr]\nonumber\\
& & =-{(\sqrt{2})^6\over3}{M_\pi^2\over{64F_\pi^6}} 
\Biggl[{{55}\over 6}L-{{173}\over{576\pi^2}}+
{{55}\over{192\pi^2}}ln{M_\pi^2\over\mu^2}
+{{53\sqrt{5}}\over{192\pi^2}}ln\biggl({{\sqrt{5}+1}\over{\sqrt{5}-1}}
\biggr)\nonumber\\
& & +{1\over{2\pi}}ln\biggl({{1+\sqrt{5}}\over2}\biggr)\Biggr]
\end{eqnarray}

(e)
\begin{eqnarray} 
\label{eq:Teloop1}
T_{[e]} & = & 
-(\sqrt{2})^8{1\over{48F_\pi^6}}\biggl[I_3
+3M_\pi^2I_6\biggr]\nonumber\\
& & ={{M_\pi^2(\sqrt{2})^8}\over{48F_\pi^6}}\biggl[{8\over3}L
-{1\over{96\pi^2}}\biggl(11 - 8 ln{M_\pi^2\over\mu^2}\biggr)
+{{7\sqrt{5}}\over{96\pi^2}}ln\biggl({{\sqrt{5}+1}\over{\sqrt{5}-1}}
\biggr)\biggr]\nonumber\\
& &  
\end{eqnarray}

(f)
\begin{eqnarray} 
\label{eq:Tfloop1}
T_{[f]} & = & 2{{(\sqrt{2})^6}\over{96F_\pi^6}}
I_3\nonumber\\
& & ={{(\sqrt{2})^6M_\pi^2}\over{48F_\pi^6}}\biggl[-{7\over6}L
-{1\over{576\pi^2}}\biggl(-29+21 ln{M_\pi^2\over\mu^2}\biggr)\nonumber\\
& &  
-{{5\sqrt{5}}\over{192\pi^2}}ln\biggl({{\sqrt{5}+1}\over{\sqrt{5}-1}}\biggr)
\biggr]
\end{eqnarray}

(g)
\begin{eqnarray} 
\label{eq:Tgloop1}
T_{[g]} & = & -{{(\sqrt{2})^6g_A^2}\over{576F_\pi^6}}
\biggl[\biggl(-{3\over2}I_8 + 9I_{10}\biggr)
\vec\sigma^{(1)}\cdot\vec P\vec\sigma^{(2)}\cdot\vec P
\nonumber\\
& & -9I_9\vec\sigma^{(1)}\cdot\vec\sigma^{(2)}
\biggr]\ .
\end{eqnarray}

Using 
\begin{equation} 
\label{eq:I910}
I_{10} 
= {1\over 3}I_8;\ 
I_9 = -{1\over 2}\Delta_\pi(0, M_\pi^2),
\end{equation}
one gets:
\begin{eqnarray}
\label{eq:Tgloop2}
T_{[g]} & = & -{{(\sqrt{2})^6g_A^2}\over{576F_\pi^6}}\biggl[\biggl(-3L
-{3\over{32\pi^2}}(1+ln{M_\pi^2\over\mu^2})\biggr)
\vec\sigma^{(1)}\cdot\vec P 
\vec\sigma^{(2)}\cdot\vec P\nonumber\\
& & +\biggl(9M_\pi^2L + {{9M_\pi^2}\over{32\pi^2}}
ln{M_\pi^2\over\mu^2}\biggr)
\vec\sigma^{(1)}\cdot\vec\sigma^{(2)}\biggr].
\end{eqnarray}

One thus gets:
\begin{equation} 
\label{eq:Tgloop3}
\langle S=0|T_{[h]}|S=0\rangle=
-{{(\sqrt{2})^6g_A^2}\over{576F_\pi^2}}\biggl[-24L
+{1\over{32\pi^2}}\biggl(3-24 ln{M_\pi^2\over\mu^2}\biggr)\biggr].
\end{equation}

(h)
\begin{eqnarray} 
\label{eq:Thloop1}
T_{[h]} & = & 
2\times{{2(\sqrt{2})^6g_A^2}\over{384F_\pi^6}}\biggl[
\biggl(- 6I_9 -  6M_\pi I_{13} \biggr)
\vec\sigma^{(1)}\cdot\vec\sigma^{(2)}\nonumber\\
& & +\biggl( -I_8-2 M_\pi I_{12} 
+ 6I_{10} + 6M_\pi I_{14}\biggr)\vec\sigma^{(1)}\cdot\vec P 
\vec\sigma^{(2)}\cdot\vec P\biggr]\nonumber\\
& & = 2\times{{2(\sqrt{2})^6g_A^2}\over{384F_\pi^6}}\biggl[\biggl(- 6 L
-{3\over{16\pi^2}}
ln{M_\pi^2\over\mu^2}+{1\over{4\pi^2}}[3 - {{3\pi^2}\over8}]
\biggr)M_\pi^2
\vec\sigma^{(1)}\cdot\vec\sigma^{(2)}\nonumber\\
& & +\biggl(-2L-{3\over{16\pi^2}}-{1\over{16\pi^2}}
ln{M_\pi^2\over\mu^2}+{3\over{32}}\biggr)
\vec\sigma^{(1)}\cdot\vec P\vec
\sigma^{(2)}\cdot\vec P\biggr].
\end{eqnarray}

One thus gets:
\begin{eqnarray}
\label{eq;Thloop2}
 \langle S=0|T_{[h]}| S=0\rangle
& = & 
(\sqrt{2})^6{{g_A^2M_\pi^2}\over{96F_\pi^6}}
\biggl[20L-{1\over{16\pi^2}}\biggl(
33-3\pi^2\biggr)+{5\over{8\pi^2}}
ln{M_\pi^2\over\mu^2}\biggr].\nonumber\\
& & 
\end{eqnarray}

Finally, the total 1-loop contribution to the DCX amplitude
is given by :
\begin{eqnarray}
\label{eq:Ttotalloop}
 & & \langle S=0|T_{total}(1-loop)|S=0\rangle \nonumber\\
& = & {{(\sqrt{2})^6M_\pi^2}\over{24F_\pi^6}}\Biggl(
\biggl[
4g_A^2+{{15}\over{16}}\biggr]L+
\biggl[-{{181}\over{256\pi^2}}+{3\over{64}}\biggr]g_A^2
-{{157}\over{1536\pi^2}}\nonumber\\
& & +{1\over{64\pi^2}}ln{M_\pi^2\over\mu^2}
\biggl[{15\over8}+5g_A^2\biggr]
+{\sqrt{5}\over{24\pi^2}}
ln\biggl({{\sqrt{5}+1}\over{\sqrt{5}-1}}\biggr)\biggl[-{{11}\over{10}}
g_A^2+{{39}\over{64}}\biggr]\nonumber\\
& & -{1\over{16\pi}}ln\biggl[{{1+\sqrt{5}}\over2}\biggr]\Biggr)
\end{eqnarray}

The numerical values of the UV-finite parts of the 
1-loop amplitudes and their UV-divergent 
portions, are given in Table II.

\clearpage

\begin{table}[htbp]
\centering
\caption{Amplitudes for 1-loop graphs}
\begin{tabular} {|c|c|c|} \hline
$\langle S=0|T_{[x]}|S=0\rangle$ & 
UV-Finite Integral (units of MeV$^{-4}$) 
& UV-Divergent Portion \\
& &  (coef of ${{(\sqrt{2})^6M_\pi^2}\over{F_\pi^6}})L$ 
\\ \hline
a & -- & -- \\ \hline
b &  $-7.7\times10^{-10}$ & $2\times {{29g_A^2}\over{288}}$ \\ \hline
c &  $10^{-9}$ & $-2\times{{41g_A^2}\over{288}}$ \\ \hline
d &  $1\times10^{-11}$ & $-{{55}\over{1152}}$ \\ \hline
e &  $-2.9\times10^{-10}$ & ${1\over9}$ \\ \hline
f &  $7\times10^{-11}$ & $-{7\over{288}}$ \\ \hline
g &  $1.9\times10^{-10}$ & ${{g^0_A}^2\over{24}}$ \\ \hline
h &  $-10^{-9}$ & ${5{g^0_A}^2\over{24}}$ \\ \hline
total &  $-(7.9\pm 1.6)\times10^{-10}$
& ${1\over{24}}\times\biggl({{15}\over{16}}+4g_A^2\biggr)$ \\ 
& ${\rm for}\ \mu=1\pm0.5\ {\rm GeV}$ & \\
\hline
\end{tabular}
\end{table}

\clearpage

Hence, comparing the numerical
values of (\ref{eq:Ttotal}) and (\ref{eq:Ttotalloop}) one sees that 
the 1-loop graphs contribute an increase of about a 1.6$\%$
relative to the tree graphs, after removal of  divergent terms
(See Section V). This agrees with the expectation that the second-order
correction in the chiral expansion should be smaller by a factor of 
$M_\pi^2/(4\pi F_\pi)^2 \simeq 0.014$ than the leading order.
The suppression of the 2-nucleon graphs relative to the leading order tree 
graphs can be seen explicitly by comparing the expressions ({\ref{eq:Ttotal}) 
and (\ref{eq:Ttotalloop}).  (This is in contrast to the large 1-loop 
corrections for the $\pi-\pi$ and $\pi-N$ vertices, which is discussed in 
Section VI.) Also note the relative insensitivity of the total (finite) 1-loop 
amplitude to the choice of the renormalization point $\mu$, as shown in the 
last line of Table II; a 50$\%$ change in $\mu$ gives a 20$\%$ change in 
amplitude.

From Table II, one sees that the dominant 1-loop contributions
come from graphs (b), (c) and (h). This
is probably related to the following observations. 
First, as is clear from Table I, the 1-nucleon-2$\pi$-propagator loop 
integrals dominate over the 2$\pi$-propagator integrals.
\footnote{Graph (d) also has a 1-nucleon-2$\pi$-propagator
structure, but the corresponding loop integrals occur
in such a way in the amplitude that there is large cancellation between
the contributions of $I_k, k =1, 3, 11$.} 
Second, graphs (b), (c) and (h) are proportional to $g_A^2>1$, as compared
to graphs (d), (e) and (f) (which do not have a $g_A^2$ dependence).

\section{Renormalization by Contact Terms}

In this section we discuss the renormalization of the 1-loop
graphs evaluated in Section IV. To remove the UV divergence 
in 3+1 dimensional space, one has to look for terms whose 
contributions precisely cancel the coefficient of the ``L" in the (2-nucleon)
1-loop amplitudes.
Given the structure of the UV-divergent parts of the 1-loop amplitudes, 
2$\pi$ - 2nucleon contact terms 
of ${\cal O}(q^2)$ should do the job as the correct counter terms, as will
become clear from (\ref{eq:contstruc1}) and (\ref{eq:me1}) below.

The contact terms are defined by:
\begin{equation} 
\label{eq:contdef}
\lim_{x\rightarrow y}\langle
\alpha^\prime|{\bar{\rm H}}{\cal O}_1\rm H(x)|\alpha\rangle
\langle\beta^\prime|{\bar{\rm H}}{\cal O}_2\rm H(y)|\beta\rangle,
\end{equation}
with $\alpha\equiv p, I_3$. 
From the 
expressions for  the amplitudes for 
the eight 1-loop graphs in Section IV, before
setting $\vec P^2=q_1\cdot q_2=M_\pi^2$, one can
show that the UV-divergent terms have the following
structures  (omitting the overall isospin factor 
$\tau_+^{(1)}\tau_+^{(2)}$ from (\ref{eq:contstruc1}))
\begin{eqnarray} 
\label{eq:contstruc1}
& & 
\biggl(\vec P^2\
 {\rm or}\ q_1\cdot q_2\ {\rm or}\  M_\pi^2\biggr)\times
{\bf 1};\nonumber\\
& & \biggl(\vec P^2\ {\rm or}\ q_1\cdot q_2\ 
{\rm or}\ M_\pi^2\biggr)\times\vec\sigma^{(1)}\cdot\vec\sigma^{(2)};
\nonumber\\
& & \vec\sigma^{(1)}\cdot\vec P 
\vec\sigma^{(2)}\cdot\vec P .
\end{eqnarray}
On taking the Fourier transform of the amplitudes 
(\ref{eq:contstruc1}), one gets the 
local forms:
\begin{equation}
\label{eq:me1}
\sim
(M_\pi^2,\ q_1\cdot q_2,\ 
\partial_i^1\partial_j^2)\delta^{(3)}(\vec x_1-\vec x_2)
\equiv O(q^2).
\end{equation}
Thus one requires O$(q^2,\phi^2)$ 2 $\pi$-2 nucleon
contact terms with  
$\tau_+^{(1)}\tau_+^{(2)}$ isospin structure.

For the purpose of renormalization, one needs to consider nine
2$\pi$ - 2nucleon contact terms, written in terms of:
\begin{equation}
u_\mu, \stackrel{\leftarrow+\rightarrow}{\rm D_\mu}, \chi_+,\ {\rm and}\ \phi,
\end{equation}
$\biggl(\stackrel{\leftarrow+\rightarrow}{\rm D_\mu}\equiv
(\partial_\mu+\Gamma_\mu)$
$+(\stackrel{\leftarrow}{\partial}-\Gamma_\mu);\ 
\chi_+= M^2(U+U^\dagger)\biggr)$. The finite parts of 
these nine terms  will not contribute, as 
explained later in this section.

One needs to include at least $\phi$ explicitly 
as a building block for DCX for the
following reason. The $\phi$ in (A.2) can be  expanded in
terms of the generators of the nucleon isospin group as :
\begin{equation}
\label{eq:phi1}
\phi=\tau_3\phi^0+\sqrt{2}(\tau_+\phi^-+\tau_-\phi^+),
\end{equation}
where $\tau_\pm={1\over 2}(\tau_1\pm i\tau_2)$, and $\phi^+$ either
anihilates $\pi^-$ or creates $\pi^+$, and $\phi^-$ either anihilates $\pi^+$ 
or creates $\pi^-$. In DCX, a $\pi^+$ goes over to $\pi^-$, implying that 
one requires
${\cal O}_1(x)$ and ${\cal O}_2(y)$ 
in (\ref{eq:contdef}) to consist at least of $\phi_-(x)
\phi_-(y)$ (or derivatives thereof), or in terms of the isospin
generators, $\tau^{(1)}_+\tau^{(2)}_+$,
where the superscripts refer to nucleons 1 and 2. 
Further, using Table III:
\clearpage
\begin{table}[htbp]
\centering
\caption{Properties of Building Blocks}
\begin{tabular}{|c|c|c|} \hline
Building Block & Isospin nature & Chiral Order \\ \hline
$\rm D_\mu$ & isoscalar($\equiv\partial_\mu$) & O($q$) \\
&  +isovector($\equiv\Gamma_\mu$) & \\ \hline
$u_\mu$   & isovector & O$(q)$ \\ \hline
$\chi_+$ & isoscalar & O$(q^2)$ \\ \hline
$\phi$ & isovector & O(1)   \\ \hline
\end{tabular}
\end{table} 

\noindent one can construct Table IV for ${\cal O}_1(x){\cal O}_2(y)$
relevant to DCX (again omitting $\tau_+^{(1)}\tau_+^{(2)}$) 
which has to include terms of the type isovector$\times$isovector.
\begin{table} [htbp]
\centering
\caption{${\cal O}_1(x){\cal O}_2(y)$ relevant to DCX}
\begin{tabular} {|c|c|} \hline
Momentum-spin form & Coordinate-spin operators 
 \\ \hline
$M_\pi^2\times{\bf 1}$ & $\phi(x)\phi(y)\langle\chi_+(x)\rangle$
\\ 
$q_1\cdot q_2\times{\bf 1}$ & $u_\mu(x)u^\mu(y);\ 
v\cdot u(x)v\cdot u(y)$ \\ 
$\vec P^2\times{\bf 1}$ & $\phi(x)
\stackrel{\leftarrow+\rightarrow}{\rm D_\mu}(x)\phi(y)
\stackrel{\leftarrow+\rightarrow}{\rm D^\mu}(y)$ \\ \hline
$M_\pi^2\times\vec\sigma^{(1)}\cdot\vec\sigma^{(2)}$ & ${\rm S}^{(1)}_\mu
\phi(x){\rm S}^{(2),\mu}\phi(y)\langle\chi_+\rangle$ \\
$q_1\cdot q_2\times\vec\sigma^{(1)}\cdot\vec\sigma^{(2)}$ &  $u_\mu(x)
{\rm S}_\nu^{(1)}u^\mu{\rm S}^{\nu,(2)};\ v\cdot u(x){\rm S}^{\mu,(1)} 
v\cdot u(y){\rm S}^{(2)}_\mu$\\
$\vec P^2\times\vec\sigma^{(1)}\cdot\vec\sigma^{(2)}$ 
& $\phi(x){\rm S}^{(1),\mu}
\stackrel{\leftarrow+\rightarrow}{\rm D_\nu}(x)\phi(y)
{\rm S}^{(2)}_\mu\stackrel{\leftarrow+\rightarrow}{\rm D^\nu}(y)$
\\ \hline
$\vec\sigma^{(1)}\cdot\vec P\vec\sigma^{(2)}\cdot\vec P 
$ & $\phi(x){\rm S}^{(1)}\cdot
\stackrel{\leftarrow+\rightarrow}{\rm D}(x)\phi(y){\rm S}^{(2)}\cdot
\stackrel{\leftarrow+\rightarrow}{\rm D}(y)$ \\ \hline
\end{tabular}
\end{table}
\clearpage

Note that ${\rm S}\cdot u(x){\rm S}\cdot u(y)$ will not contribute 
in the static limit and at  threshold (use 
(\ref{eq:cm}) and $v_{1,2}^\mu=(1,\vec 0)$).
The relevant counter terms are listed below:
\begin{eqnarray} 
\label{eq:9contterms}
\lim_{x\rightarrow y} & & \biggl[{\alpha_1\over{F_\pi^4}}
{\bar{\rm H}}(x){{\phi(x)}\over{F_\pi}}\rm H(x){\bar{\rm H}}(y)
{{\phi(y)}\over{F_\pi}}\rm H(y)\langle\chi_+(x)\rangle\nonumber\\
& & +{\alpha_2\over{F_\pi^4}}{\bar{\rm H}}(x)u_\mu(x)\rm H(x){\bar{\rm H}}(y)
u^\mu(y)\rm H(y)\nonumber\\
& & {\alpha_3\over{F_\pi^4}}
{\bar{\rm H}}(x)v\cdot u(x){\rm H}(x){\bar{\rm H}}(y)v\cdot u(y){\rm H}(y)
\nonumber\\
& & {\alpha_4\over{F_\pi^4}}{\bar{\rm H}}(x){{\phi(x)}\over{F_\pi}}
\stackrel{\leftarrow+\rightarrow}{\rm D^\mu}(x){\rm H}(x){\bar{\rm H}}(y)
{{\phi(y)}\over{F_\pi}}\stackrel{\leftarrow+\rightarrow}{\rm D_\mu}(y)
{\rm H}(y)\nonumber\\
& & +{\alpha_5\over{F_\pi^4}}{\bar{\rm H}}(x){{\phi(x)}\over{F_\pi}}
\rm S_\mu^{(1)}\rm H(x){\bar{\rm H}}(y){{\phi(y)}\over{F_\pi}}\rm S^{(2),\mu}
\rm H(y)\langle\chi_+(x)\rangle\nonumber\\
& & +{\alpha_6\over{F_\pi^4}}{\bar{\rm H}}(x)u_\mu(x)\rm S^{(1)}_\nu\rm H(x)
{\bar{\rm H}}(y)u^\mu(y)\rm S^{(2),\nu}\rm H(y)\nonumber\\
& & +{\alpha_7\over{F_\pi^4}}{\bar{\rm H}}(x)v\cdot  u{\rm S}^{\mu,(1)}
{\rm H}(x){\bar{\rm H}}(y)v\cdot u(y){\rm S}^{(2)}_\mu{\rm H}(y)\nonumber\\
& & +{\alpha_8\over{F_\pi^4}}{\bar{\rm H}}(x){\rm S}^{(1)}_\mu{{\phi(x)}
\over{F_\pi}}\stackrel{\leftarrow+\rightarrow}{\rm D^\nu}(x){\rm H}(x)
{\bar{\rm H}}(y){\rm S}^{(2),\mu}{{\phi(y)}\over{F_\pi}}
\stackrel{\leftarrow+\rightarrow}{\rm D_\nu}(y){\rm H}(y)\nonumber\\ 
& & +{\alpha_9\over{F_\pi^4}}{\bar{\rm H}}(x){{\phi(x)}\over{F_\pi}}
\rm S^{(1)}\cdot\stackrel{\leftarrow+\rightarrow}{\rm D}(x)\rm H(x)
{\bar{\rm H}}(y){{\phi(y)}\over{F_\pi}}
\rm S^{(2)}\cdot\stackrel{\leftarrow+\rightarrow}{\rm D}(y)\rm H(y)\biggr].
\nonumber\\
& & 
\end{eqnarray}
The coupling constants $\{\alpha_i\}$ have been made dimensionless by
construction of the terms.

It  should be noted that the  following
five types of UV-finite 2$\pi$-2nucleon
O$(q^2)$ contact terms  will  also contribute to off-threshold DCX:
\begin{eqnarray} 
\label{eq:5contterms}
& & \lim_{x\rightarrow y}\biggl[
\epsilon^{\mu\nu\rho\lambda}v_\rho{\bar{\rm H}}(x)
u_\mu(x){\rm H}(x){\bar{\rm H}}(y) {\rm S}^{(2)}_\nu u_\lambda(y){\rm H}(y);
\nonumber\\
& & \epsilon^{\mu\nu\rho\lambda}v_\rho{\bar{\rm H}}(x)
u_\mu(x)
{\rm H}(x){\bar{\rm H}}(y)\phi(y){\rm S}^{(2)}_\nu{\rm D}_\lambda{\rm H}(y);
\nonumber\\
& & \epsilon^{\mu\nu\rho\lambda}v_\rho{\bar{\rm H}}(x)
{\rm S}^{(1)}_\nu u_\mu(x){\rm H}(x)
{\bar{\rm H}}(y)\phi(y){\rm D}_\lambda{\rm H}(y);
\nonumber\\
& & \epsilon^{\mu\nu\rho\lambda}v_\rho{\bar{\rm H}}(x)
u_\mu{\rm D}_\lambda
{\rm S}^{(1)}_\nu{\rm H}(x){\bar{\rm H}}(y)\phi(y){\rm H}(y);
\nonumber\\
& & \epsilon^{\mu\nu\rho\lambda}v_\rho{\bar{\rm H}}(x)
u_\mu(x){\rm D}_\lambda{\rm H}(x){\bar{\rm H}}\phi(y)
{\rm S}^{(2)}_\nu{\rm H}(y)\biggr].
\end{eqnarray}

By renormalization of loop-graph integrals, one  gets a constraint on a linear
combination of the UV-divergent parts of LEC's of the seven 
of the nine 2$\pi$ - 2nucleon contact counter terms of (\ref{eq:9contterms}).
First one calculates the amplitudes T corresponding to the contact terms of
(\ref{eq:9contterms}). Then, writing $\alpha_i=\alpha_i^r+\lambda_i L$,
(where $L$ was defined in (\ref{eq:Ldef})), and comparing with the UV-divergent
portion of the total 1-loop amplitude (\ref{eq:Ttotalloop}) and Table II, 
one gets the following conditions (suppressing
the two pairs of Pauli spinors and using
 (\ref{eq:S=0me}) - (\ref{eq:kinptdef})):

(1) spin-independent renormalization (graphs [(a)+...+(f)]):

\begin{equation} (\sqrt{2})^2\biggl(4\lambda_1+2\lambda_2-2\lambda_3\biggr)
=(\sqrt{2})^6\biggl(-{g_A^2\over{12}}-{5\over{128}}\biggr);
\end{equation}

(2) spin-dependent renormalization (graphs [(g)+(h)]):

\begin{equation}
(\sqrt{2})^2\biggl(3\lambda_5+{3\over 2}\lambda_6
+6\lambda_7-2\lambda_9\biggr)=-(\sqrt{2})^6{g_A^2\over{4}},
\end{equation}
which on addition gives:
\begin{equation} (\sqrt{2})^2\biggl(4\lambda_1+2\lambda_2-2\lambda_3+3\lambda_5
+{3\over 2}\lambda_6+6\lambda_7-2\lambda_9\biggr)
=-{{(\sqrt{2})^6}\over{24}}
\biggl({{15}\over{16}} + 4g_A^2\biggr)\ .
\end{equation}
Note that the $\lambda_4, \lambda_8$ terms do not contribute at the 
kinematic point (\ref{eq:kinptdef}). So, it is clear that the
renormalization of
the loop integrals can be accomplished by the 2$\pi$-2nucleon
contact terms.

Because the nucleon-nucleon interaction  has a strong
short range repulsion, the nucleon-nucleon wave function vanishes at short
relative distances. Since the contact terms in (\ref{eq:9contterms}) and 
(\ref{eq:5contterms}) behave as $\delta^{(3)}(\vec x-\vec y)$,
one does not get any contribution from these 
2$\pi$ - 2nucleon contact terms, as well as from the UV-divergent parts of
the 1-loop integrals. Hence, 
only UV-finite parts of the loop  integrals contribute
to  the DCX loop amplitudes after renormalization.

\section{Summary and Discussion}

The goal of this paper has been to calculate the one loop correction
to the 2-nucleon amplitudes for  pion double charge exchange (DCX) scattering
by a nuclear target, at threshold, in the framework of HBChPT. 
For a numerical estimate of the 2-nucleon-1-loop correction,
the amplitudes are evaluated at a typical kinematic point (\ref{eq:kinptdef}). 
These are compared to the leading order amplitudes (for the tree graphs),
obtained in the same theoretical framework.

The 2-nucleon-1-loop graphs give a threshold contribution of about 0.016 
relative to the leading order tree graphs
at the same kinematics [with $(\vec p^\prime-\vec p)^2=M_\pi^2$]. 
Both calculations are done in the static limit and the impulse 
approximation for the nucleons, and using the renormalized values of the
axial-vector coupling, pion-decay constants and the pion and nucleon masses. 
As remarked at the end of Section V, these corrections are of the order
of $M_\pi^2/(4\pi F_\pi)^2 \simeq 0.014$ times the tree graphs in LO 
($\nu = 0$). This is as expected for a chiral correction of $\nu = 2$.

As noted in Section II, there are also corrections to the LO tree graphs to 
order $\nu = 2$ arising from corrections to the $\pi-\pi$ and $\pi-N$ vertices,
which have been studied in the literature.
The $\pi-\pi$ vertex has been determined to the required order (to one loop)
\cite{gl}, with the LECs fixed phenomenologically. However, although the 
$\pi-N$ vertex corrections have also been studied through one loop
\cite{bkm3,bkm2}, the LECs have only been determined for the on-shell
threshold cases of $\pi-N$ scattering lengths \cite{bkm3} and $\pi^\pm$ 
production on a single nucleon \cite{bkm2}. Therefore, there is not enough 
information from these studies to fix all the vertex LECs required to carry 
out the full $\nu=2$ calculation of the DCX tree graphs. 

However, the 1-loop calculations cited do show large corrections to the
vertices. We can estimate how much this would change the DCX tree-graph 
amplitudes by using the on-shell results as follows. First, we characterize 
the three amplitudes of 
(\ref{eq:Ta}),(\ref{eq:Tb}) and (\ref{eq:Tc}) in terms of vertex coefficients:

\begin{eqnarray}
\label{eq:Tabc}
T_{[a]} & \sim & D_1 g_A, \nonumber \\
T_{[b]} & \sim & A g_A^2, \\
T_{[c]} & \sim & (a^-)^2. \nonumber
\end{eqnarray}

The amplitude $D_1$ is the $\pi N \rightarrow \pi\pi N$ threshold 
amplitude  \cite{bkm2} that contributes to the DCX tree graph [a].
To leading order, $D_1(LO) \sim g_A/F_\pi^2$. The corrections through one loop 
give \cite{bkm2}
\begin{equation}
D_1 = D_1(LO)(1 + 0.67).
\end{equation}
The loop correction is actually 0.15; the larger correction, 0.52, is
from a `recoil' term of order $M_\pi^2/m$.

The amplitude $A$ gives the DCX contribution to tree graph [b] from the 
$\pi-\pi$ vertex, and can easily be shown to be to the two s-wave scattering 
lengths $a_I$, with isospins $I =$  0, 2, by
\begin{equation}
A = (2a_0 + a_2)/3.
\end{equation}
To LO, $A(LO) \sim M_\pi/F_\pi^2$. The 1-loop corrections \cite{gl} change
this to
\begin{equation}
A = A(LO)(1 + 0.30).
\end{equation}

The isovector $\pi N$ scattering length $a^-$ contributes to the 
double-scattering
tree graph [c]. To LO, $a^-(LO) \sim M_\pi/F_\pi^2$. With corrections through 
1-loop, this becomes \cite{bkm3}
\begin{equation}
a^- = a^-(LO)(1 + 0.046).
\end{equation}
This small correction masks a 20$\%$ contribution of the loop, partially cancelled
by the `recoil' correction.

If we now use these corrected values for the vertices in the expressions (\ref{eq:Tabc}),
we obtain the following estimates of their effect to order $\nu =2$ on the the 
tree-graph amplitudes:
\begin{eqnarray}
T_{[a]} & = & 1.67\ T_{[a]}(LO) \nonumber \\
T_{[b]} & = & 1.30\ T_{[b]}(LO) \\
T_{[c]} & = & 1.092\ T_{[c]}(LO). \nonumber
\end{eqnarray}
These corrections are much larger than the 1-loop corrections of this paper,
which involve both nucleons participating in the DCX reaction. However, the
vertex corrections are only estimates, since not all the off-shell vertex
coefficients are determined.

So we are led to two conclusions about corrections to DCX to order $\nu = 2$.
First, the one-loop correction to the tree graph amplitudes is small, whether
compared to the LO tree amplitudes, or to those amplitudes also corrected to
one-loop order. The dominant contributions (graphs 3b,3c, and 4h) all have
one nucleon propagator, and are proportional to $g_A^2>1$. The largest, 4h,
is clearly a correction to the largest tree graph, 2c, for double scattering. 
(Triple scattering on two nucleons has not been included; it belongs with other
triple scattering graphs on three nucleons.) The result is weakly dependent
on the choice of the renormalization point $\mu$.

For the purpose of renormalization of the loop graphs, we found that nine 
2$\pi$ - 2nucleon contact terms are required, of which 
seven  contribute at the specific kinematic point considered (\ref{eq:kinptdef}).
However, since we have assumed that the nuclear wavefunctions 
vanish near zero (relative) distance of nucleon separation,
none of the (UV-finite parts) of the contact terms will contribute, because 
of the $\delta^{(3)}(\vec x-\vec y)$ in 
the forms of contact terms of (\ref{eq:9contterms}) 
and (\ref{eq:5contterms}). If this anticorrelation 
assumption were relaxed, then one would have nine 
unknown LECs for (\ref{eq:9contterms})
(seven for the kinematic point (\ref{eq:kinptdef})), 
and five for (\ref{eq:5contterms}), to be determined 
empirically, to the order considered [O$(q^2)$].

The second conclusion is that the one-loop corrections to the
$\pi-\pi$ and $\pi-N$ vertices produce the largest corrections at order
$\nu = 2$ to the LO tree amplitudes. The large correction to the $\pi-\pi$
vertex has been related to the pion radius \cite{gl}, that is, to the
correction for the LO assumption of a point pion. Presumably, similar
effects, including resonant contributions, give the large $\pi-N$ vertex
corrections. It is clear that these corrections are anomalously large,
while the one-loop corrections to the two-nucleon amplitudes are of
expected size for the O$(q^2)$ terms. The latter may reflect the fact
that the $N-N$ system is already of finite size.

However, as noted above, the vertex corrections for the $\pi-N$ amplitudes
have not been fixed sufficiently for a  complete calculation of the DCX tree
amplitudes. This would require an extension of the 1-loop theory of the
various $\pi-N$ amplitudes to off-shell kinematics, beyond what has
already been done in the literature. 
As a practical matter, the $\pi-N$ on-shell constants could be taken directly
from experiment, as in the impulse approximation, rather than from higher
orders in chiral perturbation theory.

\section*{Acknowledgements}

We would like to thank N. Kaiser and Ulf-G.Meissner for
their continuous help, in terms of clarifications and preprints.
This research was supported in part by the U.S.
Department of Energy under 
Grant No. DE-FG02-88ER40425 with the University of Rochester.

\section*{Appendix}
\appendix
\section{Vertices and $f$'s for 1-Loop Graphs for $\pi$-NN}
\setcounter{equation}{0}
\seceqaa

The pion field is represented as a matrix-valued SU(2) field $U$ defined 
as following:
\begin{equation} e^{i{\phi\over{F_\pi}}}\equiv U(\pi),
\end{equation}
where $\phi$ is a traceless (as det U=1) matrix written  in terms of the pion
triplet as:
\begin{equation}\sqrt{2}\left( \begin{array} {cc}
{\pi^0\over \sqrt{2}} & \pi^+ \\
\pi^- & -{\pi^0\over \sqrt{2}}
\end{array} \right).
\end{equation}
The matrix is written in  a space that corresponds
to the standard ($p/n$) isospinors
which  are implied in  the H 
notation.

\subsection{Vertices}

The leading order terms in (H)ChPT, written using (A1) and (A2),
that are used for the evaluation of ${\rm m}\pi-{\bar{\rm N}}{\rm N} 
(m=1,2,3,4)$ and 4-$\pi$
vertices, are written out.

\begin{eqnarray} & & 
1\pi{\bar{\rm N}}{\rm N}:g_A{\bar{\rm H}}{\rm S}\cdot u^{(1)}\rm H
\nonumber\\
& &= -(\sqrt{2}) {g_A\over F_\pi}\biggl[{1\over\sqrt{2}}{\bar{\rm p}}{\rm S}
\cdot\partial\pi^0
{\rm p}+{\bar{\rm p}}{\rm S}\cdot\partial\pi^+{\rm n}
+{\bar{\rm n}}{\rm S}
\cdot\partial\pi^-\rm n
-{1\over{\sqrt{2}}}{\bar{\rm n}}{\rm S}\cdot\partial\pi^0\rm n\biggr]
\nonumber\\
& & \nonumber\\
& & 2\pi{\bar{\rm N}}{\rm N}:i{\bar{\rm H}}
v\cdot\rm D^{(2)}\rm H\equiv{\bar{\rm H}}v\cdot\Gamma^{(2)}\rm H\nonumber\\
& & =(\sqrt{2})^2{i\over{8F^2_\pi}}\biggl[{\bar{\rm p}}\rm p\pi^{[+}v\cdot\partial\pi^{-]}
+\sqrt{2}{\bar{\rm p}}\rm n\pi^{[0}v\cdot\partial\pi^{+]}+\sqrt{2}{\bar{\rm n}}
\rm p\pi^{[-}v\cdot\partial\pi^{0]}+{\bar{\rm n}}\rm n\pi^{[-}v\cdot\partial
\pi^{+]}\biggr]\nonumber\\
& & \nonumber\\
& & 3\pi{\bar{\rm N}}{\rm N}: g_A{\bar{\rm H}}{\rm S}\cdot u^{(3)}{\rm H} 
\nonumber\\
& & =(\sqrt{2})^3{{g_A^2}\over{12F^3_\pi}}\biggl[{1\over\sqrt{2}}{\bar{\rm p}}
\biggl(2\pi^+\pi^-\rm S\cdot\partial\cdot\pi^0
-\pi^0(\pi^+\rm S\cdot\partial\pi^-+\pi^-\rm S\cdot\partial\pi^+)\biggr)
\rm p\nonumber\\
& & +{\bar{\rm p}}\biggl({\pi^0}^2\rm S\cdot\partial\pi^+-\pi^+\pi^0\rm S
\cdot\partial\pi^0+\pi^+\pi^{[-}\rm S\cdot\partial\pi^{+]}\biggr)n+\rm h.c.
\nonumber\\
& & +{1\over\sqrt{2}}{\bar{\rm n}}\biggl(-2\pi^+\pi^-\rm S\cdot\partial\pi^0
+\pi^0+\pi^0(\pi^-\rm S\cdot\partial\pi^++\pi^+\rm S\cdot\partial\pi^-)\biggr)
\rm n\biggr]\nonumber\\
& & \nonumber\\
& & 4\pi{\bar{\rm N}}{\rm N}: {\bar{\rm H}}iv\cdot\rm D^{(4)}\rm H
\equiv{\bar{\rm H}}iv\cdot\Gamma^{(4)}\rm H\nonumber\\
& &(\sqrt{2})^4 {i\over{96F^4_\pi}}\biggl({{\pi^0}^2\over2}+\pi^+\pi^-\biggr)
\biggl[{\bar{\rm p}}\rm p\pi^{[-}v\cdot\partial\pi^{+]}+(\sqrt{2}
{\bar{\rm p}}\rm n\pi^{[+}v\cdot\partial\pi^{0]}+\rm h.c.)\nonumber\\
& & +{\bar{\rm n}}{\rm n}\pi^{[+}v\cdot\partial\pi^{-]}
\biggr]\nonumber\\
& & \nonumber\\
& & 4\pi:{F^2_\pi\over4}\biggl(\langle\partial^\mu U^\dagger\partial_\mu 
U\rangle+M_\pi^2\langle(U^\dagger+U-2)\rangle\biggr)^{(4)}\nonumber\\
& & =
(\sqrt{2})^4{1\over{24F^2_\pi}}\biggl[-2\pi^+\pi^-\partial_\mu\pi^+\partial^\mu\pi^-+(\pi^-)^2
(\partial_\mu\pi^+)^2+(\pi^+)^2(\partial_\mu\pi^-)^2+
M_\pi^2(\pi^+)^2(\pi^-)^2\biggr]\nonumber\\
& & 
\end{eqnarray}

\subsection{Combinatoric Factors for 1-Loop Graphs}

The combinatoric factors ``f" of (\ref{eq:Sndef})
  for 1-loop graphs, are evaluated.
First, $S^{(n)}$ (See (\ref{eq:Sndef})) is
written out in terms of ``${\bar{\rm H}}\phi^m\rm H$," $m=1, 2, 3, 4$ and
``$\phi^4$," which represent the $m\pi-n_i$ 
nucleon vertices  with
$m=1, 2, 3, 4; n_i=2$ and $m=4; n_i=0$. Then, using  the time ordering
properties of bosonic and fermionic bilinear fields, those
terms relevant to the DCX 1-loop graphs are picked out.
Then on comparison with (\ref{eq:Sndef}), 
the combinatoric factors are read off for  the
eight 2-nucleon 1-loop graphs. They are listed along with the
corresponding forms of equation (\ref{eq:Sndef}) in Table V.
 
\clearpage

\begin{table}[htbp]
\centering
\caption{Combinatoric Factors for the 1-Loop Graphs}
\begin{tabular} {|c|c|c|c|} \hline
$x$ &  
$S^{(n)}_{[x]} = {i^n\over{n!}}\int\prod_{k=1}^n d^4x_k$ &
$fi^n\int\prod_{k=1}^n d^4x_k$ & $f$ \\ 
& ${\cal T}\biggl(\prod_{i=1}^n\sum_j {\cal L}^I_j(x_i)\biggr)$ &
${\cal T}\biggl(\prod_{i=1}^n{\cal L}^I_i(x_i)\biggr)$ & \\
& & relevant to $x$ & \\ \hline
$a$ & -- & -- & -- \\ \hline
$b$ &
${i^3\over6}\int\prod_{i=1}^3d^4x_i{\cal T}\biggl[\prod_{i=1}^3$
& $i^3\prod_{i=1}^3\int d^4x_i$ & 1 \\
& $\biggl({\bar{\rm H}}(\phi+\phi^2+\phi^3)\rm H\biggr)(x_i)$
& ${\cal T}\biggl[\biggl({\bar{\rm H}}\phi
\rm H\biggr)(x_1)\biggl({\bar{\rm H}}\phi^2\rm H\biggr)(x_2)$ & \\
& & $\biggl({\bar{\rm H}}\phi^3\rm H\biggr)(x_3)\biggr]$ & \\ \hline
$c$ & ${i^4\over 24}\prod_{i=1}^4\int d^4x_i
{\cal T}\biggl[\prod_{i=1}^4\biggl({\bar{\rm H}}(\phi+\phi^2){\rm H}$
& ${1\over 2}\prod_{i=1}^4 \int d^4x_i T\biggl[\biggl({\bar{\rm H}}\phi{\rm H}
\biggr)(x_1)\biggl({\bar{\rm H}}\phi\rm H\biggr)(x_2)$ & ${1\over 2}$ \\  
& $+\phi^4\biggr)(x_i)\biggr]$ & 
$\biggl({\bar{\rm H}}\phi^2\rm H\biggr)(x_3)\phi^4(x_4)\biggr]$ & \\ \hline
$d$ &
${i^3\over6}\prod_{i=1}^3\int d^4x_i
{\cal T}\biggl[\biggl({\bar{\rm H}}\phi^2\rm H\biggr)(x_1)$ &
${i^3\over6}\prod_{i=1}^3\int d^4x_i
{\cal T}\biggl[\biggl({\bar{\rm H}}\phi^2\rm H\biggr)(x_1)$ & ${1\over 6}$ \\
& $\biggl({\bar{\rm H}}
\phi^2\rm H\biggr)(x_2)\biggl({\bar{\rm H}}\phi^2\rm H\biggr)(x_3)\biggr]$
& $\biggl({\bar{\rm H}}
\phi^2\rm H\biggr)(x_2)\biggl({\bar{\rm H}}\phi^2\rm H\biggr)(x_3)\biggr]$
& \\ \hline
$e$ & ${i^3\over 6}\int \prod_{i=1}^3d^4x_i
{\cal T}\biggl[\prod_{i=1}^3\biggl(
{\bar{\rm H}}\phi^2\rm H+\phi^4\biggr)(x_i)\biggr]$
& ${i^3\over 2}\prod_{i=1}^3\int d^4x_i{\cal T}\biggl[\biggl({\bar{\rm H}}
\phi^2\rm H\biggr)(x_1)$ & ${1\over 2}$ \\
& & $\biggl({\bar{\rm H}}\phi^2\rm H\biggr)(x_2)
\phi^4(x_3)\biggr]$ & \\ \hline
$f$ & ${i^2\over2}\prod_{i=1}^2\int d^4x_i
{\cal T}\biggl[\biggl({\bar{\rm H}}(\phi^2+\phi^4)\rm H\biggr)(x_1)$
& $i^2\prod_{i=1}^4\int d^4x_i 
{\cal T}\biggl[\biggl({\bar{\rm H}}\phi^2\rm H\biggr)(x_1)$ & 1 \\
& $\biggl({\bar{\rm H}}(\phi^2+\phi^4)\rm H\biggr)(x_2)\biggr]$
& $\biggl({\bar{\rm H}}\phi^4\rm H\biggr)(x_2)\biggr]$ & \\ \hline
$g$ &
${i^2\over 2}\prod_{i=1}^2\int d^4x_i
{\cal T}\biggl[\biggl({\bar{\rm H}}\phi^3\rm H\biggr)(x_1)$ & 
${i^2\over 2}\prod_{i=1}^2\int d^4x_i
{\cal T}\biggl[\biggl({\bar{\rm H}}\phi^3\rm H\biggr)(x_1)$ & 
${1\over2}$ \\
& $\biggl({\bar{\rm H}}\phi^3\rm H\biggr)(x_2)\biggr]$
& $\biggl({\bar{\rm H}}\phi^3\rm H\biggr)(x_2)\biggr]$
& \\ \hline
$h$ &
${i^3\over6}\int\prod_{i=1}^3d^4x_i
{\cal T}\biggl[\prod_{i=1}^3\biggl($
& $i^3\int d^4x_1\int d^4x_2\int d^4x_3 T\biggl[\biggl({\bar{\rm H}}\phi
\rm H\biggr)(x_1)$ & 1 \\
& ${\bar{\rm H}}(\phi+\phi^2+\phi^3)\rm H
\biggr)(x_i)$ & $\biggl({\bar{\rm H}}\phi^2\rm H\biggr)
(x_2)\biggl({\bar{\rm H}}\phi^3\rm H\biggr)(x_3)\biggr]$ & \\ \hline
\end{tabular}
\end{table}

\clearpage

\section{1-Loop Integrals}
\setcounter{equation}{0}
\seceqbb

There are 11 1-loop integrals defined and evaluated in this
appendix. The integrals $\gamma_{4,6}$ do not occur in the 1-loop DCX 
amplitudes
as discussed later. The notations used, though similar to the 
ones used by \cite {bkm1} and  \cite {gss}, are slightly 
different.

Four of those integrals are referred to as basic integrals
: $\Delta_\pi(0, M^2)$, $J^{\pi\rm N}$, ${\cal J}^{\pi\pi},\gamma_0$,
evaluated at different kinematic points. 
(The 0 in $\Delta_\pi(0, M^2)$ implies that
the ``external" 4-momentum squared 
``${\cal P}^2$" that appears in other integrals, 
is zero)  They are defined below. 
\begin{eqnarray} & & {1\over i}\int{{d^dk}\over{(2\pi)^d}}
{1\over{M^2-k^2-i\epsilon}}\equiv\Delta_\pi(0, M^2)\nonumber\\
& & {1\over i}\int{{d^dk}\over{(2\pi)^d}}{1\over{(v\cdot k-\omega-i\epsilon)
(M_\pi^2-k^2-i\epsilon)}}\equiv J^{\pi N}(\omega)\nonumber\\
& & {1\over i}\int{{d^dk}\over{(2\pi)^d)}}
{1\over{(M^2-k^2-i\epsilon)(M_\pi^2-(k-{\cal P})^2-i\epsilon)}}\equiv 
{\cal J}^{\pi\pi}({\cal P}^2; M^2)\nonumber\\
& & {1\over i}\int{{d^dk}\over{(2\pi)^d}}
{1\over{(v\cdot k-\omega-i\epsilon)
(M^2-k^2-i\epsilon)(M^2-(k-{\cal P})^2-
i\epsilon)}}\nonumber\\
& & \equiv\gamma_0(\omega,\Omega,{\cal P}^2),
\end{eqnarray}
with $\Omega\equiv v\cdot {\cal P}$. In \cite{gss}, ${\cal P}^2$ that 
figures in ${\cal J}^{\pi\pi}({\cal P}^2,M^2)$ is $\geq 0$.
In this paper, one requires ${\cal P}^2\leq 0$.

The integrals $\Delta_\pi(0, M^2), {\cal J}^{\pi\pi}({\cal P}^2; M^2)$ 
and $J^{\pi N}(\omega)$ 
have already been evaluated in the literature (\cite {bkm1,gss}), 
where they were denoted by $\Delta_\pi(0),
J(s\ {\rm or}\ t)$ and $J_0(\omega)$, respectively.
We have alterred the notation for $\Delta_\pi$ and $J$
to introduce $M^2$ as a variable, because we need to differentiate
these expressions with respect to $M^2$, as in (B11) and (B20).
The form of $\gamma_0(\omega,\Omega,{\cal P}^2)$, 
required for DCX, has not been evaluated previously.
It is a generalization of the integral denoted  by $\gamma_0(\omega)$
in \cite{bkm1}, for which $\omega=0,
{\cal P}^2=0,\ {\cal P}\rightarrow-{\cal P}$ ($\Omega$ used in this 
paper is denoted by ``$\omega$" in \cite{bkm1}).
The reason for introducing $\Omega$ and
${\cal P}^2$ in addition to $\omega$ in the argument of $\gamma_0$ 
becomes clear from the points (a) and (b) in the paragraph after (B3).
Also, note that the indexing ($i=1,2,..$) of ${\cal J}^{\pi\pi}_i({\cal P}^2;
M^2)$ and $\gamma_i(\omega,\Omega,{\cal P}^2)$ in (\ref{eq:appB2}) and 
(\ref{eq:appB3}) below differs from that used for the corresponding
integrals $J_{2i}^{\pi\pi}(s\ {\rm or}\ t)$ and 
$\gamma_i(\omega)$ in \cite{bkm1,gss}.

The above four
integrals are referred to as ``basic" because 
the integrands have no momentum dependence in their numerators, and are the 
most basic 
of 1$\pi$ propagator,  1-nucleon-1$\pi$ propagators, 2$\pi$ propagators and 
1-nucleon-2$\pi$ propagators integrals respectively. 

The remaining integrals have momentum dependence in numerators
(and hence have a tensorial  character), and are defined as:
\begin{eqnarray}
\label{eq:appB2}
& & {1\over i}\int{{d^dk}\over{(2\pi)^d)}}{{k_\mu\ {\rm or}\ k_\mu k_\nu}
\over{(M^2-k^2-i\epsilon)(M_\pi^2-(k-{\cal P})^2-i\epsilon)}}\equiv 
{\cal P}_\mu{\cal J}^{\pi\pi}_1({\cal P}^2; M^2);\nonumber\\
& & {\rm or}\ g_{\mu\nu}
{\cal J}^{\pi\pi}_2({\cal P}^2; M^2)+{\cal P}_\mu {\cal P}_\nu
{\cal J}^{\pi\pi}_3({\cal P}^2; M^2)
\end{eqnarray}

\begin{eqnarray}
\label{eq:appB3}
& & {1\over i}\int{{d^dk}\over{(2\pi)^d}}{{k_\mu\ {\rm or}\ k_\mu k_\nu}
\over{(v\cdot k-\omega-i\epsilon)( M^2-k^2-i\epsilon)(M^2-(k-{\cal P})^2-
i\epsilon)}}\nonumber\\
& & \equiv
v_\mu\gamma_1(\omega,\Omega,{\cal P}^2)
+{\cal P}_\mu\gamma_2(\omega,\Omega,{\cal P}^2);\nonumber\\
& & {\rm or}\ g_{\mu\nu}\gamma_3(\omega,\Omega,{\cal P}^2)+v_\mu v_\nu
\gamma_4(\omega,\Omega, {\cal P}^2)+{\cal P}_\mu {\cal P}_\nu
\gamma_5(\omega,\Omega,{\cal P}^2)\nonumber\\
& & +(v_\mu {\cal P}_\nu
+v_\nu {\cal P}_\mu)\gamma_6(\omega,\Omega,{\cal P}^2),
\end{eqnarray}
where alternative numerators in $k_\mu$ or 
$k_\mu k_\nu$ have been shown in each case.
                          
Other than notational differences, there are also
the following kinematical differences between the the integrals
of (B3) and those figuring in \cite{bkm1} 
: (a) Our ${\cal P}^2$ does not vanish, in  general, unlike the case for 
\cite{bkm1}, in which ${\cal P}^2$ is the four-momentum squared of an on-shell 
photon; (b)$\omega\neq0;\Omega\equiv v\cdot{\cal P}\neq\omega$;
(c) we consider $(k-{\cal P})^2$ instead of
$(k+{\cal P})^2$ in the integrals. 
 
Of the nine, two do not  contribute to DCX: $\gamma_{4,6}$. The reason
is that because these two integrals can occur only in 
graphs (g) and (h)
with integrands having numerators of the type
${\rm S}^{(1)}\cdot k{\rm S}^{(2)}\cdot k$, their coefficients will have
at least one ${\rm S}^{(1),(2)}\cdot v$, which vanishes in the static limit.
(That is because  in the  static limit $v^\mu=(1,\vec 0)$ and
${\rm S}^\mu=(0,{\vec\sigma\over 2}$))
The remaining seven can be rewritten as linear combinations of the four 
basic integrals of (B1).

All the integrals for the calculation
are real, and hence so are all the 1-loop amplitudes. This is because
the elastic scattering
amplitude becomes real at threshold.

Two further simplifications of the calculation are obtained by the following.

(A)

The 3$\pi$-propagator integrals that occur in 1-loop graph Fig 7.2e
using (\ref{eq:cm}), are of the type :

\begin{equation} {1\over i}\int {{d^dk}\over{(2\pi)^d}}{{f(k_\mu)}\over
{(M_\pi^2-k^2-i\epsilon)^2(M_\pi^2-(k-{\cal P})^2-i\epsilon)}},
\end{equation}

which can rewritten as :
\begin{equation} -{\partial\over{\partial M^2}}
{1\over i}\int {{d^dk}\over{(2\pi)^d}}{{f(k_\mu)}
\over{(M^2-k^2-i\epsilon)(M_\pi^2-(k-{\cal P})^2-i\epsilon)}},
\end{equation}
where $M^2$ is set equal to $M_\pi^2$ after differentiation.

(B)

Use is also made of some identities involving the loop integrals,  which are 
valid only in the static limit and impulse approximation (\ref{eq:cm}):
\begin{eqnarray} & & \gamma_0(\omega=\pm M_\pi, 
\Omega=\pm M_\pi, {\cal P}^2=(p_4-p_2\pm q_{1,2})^2)\nonumber\\
& & =\gamma_0(\omega=0, \Omega=\mp M_\pi, {\cal P}^2=(p_2-p_4
\mp q_{1,2})^2)\nonumber\\
& & =\gamma_0(\omega=0, \Omega=\mp M_\pi, {\cal P}^2=(p_4-p_2
\mp q_{1,2})^2)\nonumber\\
& & =\gamma_0(\omega=\pm M_\pi, \Omega=\pm M_\pi, 
{\cal P}^2=(p_2-p_4\pm q_{1,2})^2),
\end{eqnarray}
where $p_2^\mu\approx(0,-\vec p)$, $p_4^\mu\approx(0,-\vec p^\prime)$,
$v^\mu=(1,\vec 0)$, $v\cdot\rm S^{1,2}=0$.

Also we uses the following the identity:
\begin{eqnarray}
& & \gamma_0(\omega=\pm M_\pi,\Omega=\pm M_\pi,{\cal P}^2= M_\pi^2
-\vec P^2)\nonumber\\
& & =\gamma_0(\omega=0,\Omega=\mp M_\pi,{\cal P}^2= M_\pi^2
-\vec P^2).
\end{eqnarray}

\subsection{Basic Integrals}

The results after evaluation of the basic integrals of (B1) are given
below.

\begin{eqnarray} & & \Delta_\pi(0, 
M^2)=2 M^2 L + {M^2\over{16\pi^2}}
ln{M^2\over\mu^2}+{\cal O}(d-4);\nonumber\\
& & J^{\pi N}(\omega)=-4L\omega +{\omega\over{8\pi^2}}
(1-ln{M_\pi^2\over\mu^2})-{1\over{4\pi^2}}\sqrt{M_\pi^2-\omega^2}
\rm arccos\biggl(-{\omega\over M_\pi}\biggr)+{\cal O}(d-4);\nonumber\\
& & \theta(-{\cal P}^2){\cal J}^{\pi\pi}({\cal P}^2; M^2)=
-2L-{1\over{16\pi^2}}\biggl(-1+ln{M^2\over\mu^2}-
{{(\Delta-{\cal P}^2)}\over{2{\cal P}^2}}
ln{M^2\over M_\pi^2}\nonumber\\
& & +{{\Lambda^{1\over2}({\cal P}^2; M^2, M_\pi^2)}
\over{-2{\cal P}^2}}ln\biggl[
{{(\Lambda^{1\over2}-{\cal P}^2)^2-\Delta^2}
\over{(\Lambda^{1\over2}+{\cal P}^2)^2-\Delta^2)}}
\biggr]\biggr);\nonumber\\
& & (\Delta\equiv M^2- M_\pi^2;\  
\Lambda({\cal P}^2; M^2, M_\pi^2)\equiv
[{\cal P}^2-( M- M_\pi)^2][{\cal P}^2-( M+ M_\pi)^2])
\end{eqnarray}
\begin{equation}
I_1\equiv{\cal J}^{\pi\pi}(- M_\pi^2, M_\pi^2) 
= -2L
- {1\over{16\pi^2}}\Biggl[-1+ln{M_\pi^2\over\mu^2}+\sqrt{5}\ 
ln\biggl({{\sqrt{5}+1}\over{\sqrt{5}-1}}\biggr)\Biggr]
\end{equation}
\begin{equation}
I_8\equiv{\cal J}^{\pi\pi}({\cal P}^2=0; M^2= M_\pi^2)= -2L
-{1\over{16\pi^2}}(1+ln{ M_\pi^2\over\mu^2})+{\cal O}(d-4)
\end{equation}
\begin{eqnarray}
& & \theta(-P^2){\partial\over{\partial M^2}}{\cal J}^{\pi\pi}
({\cal P}^2; M^2, M_\pi^2)|_{M^2= M_\pi^2}
\nonumber\\
& & =-{1\over{16\pi^2}}
\biggl({1\over{M_\pi^2}}
+{1\over\sqrt{(-{\cal P}^2)
(4 M_\pi^2-{\cal P}^2)}}ln\biggl[{{\sqrt{4M_\pi^2-{\cal P}^2}+
\sqrt{-{\cal P}^2}}\over{\sqrt{4 M_\pi^2-{\cal P}^2}-\sqrt{-{\cal P}^2}}}
\biggr]\biggr);\nonumber\\
& & 
\end{eqnarray}

\begin{equation}
(i)\ \gamma_0(\omega=0,\Omega= M_\pi,{\cal P}^2=0)-\gamma_0(\omega=0,
\Omega=- M_\pi,{\cal P}^2=0)=-{1\over{32 M_\pi}};
\end{equation}
\begin{eqnarray}
& & (ii)\ I_{11}\equiv 
\gamma_0(\omega=- M_\pi,\Omega=0,{\cal P}^2=- M_\pi^2)\nonumber\\
& & = {1\over{8\sqrt{2}\pi M_\pi}}
\biggl[ln\biggl(1+\sqrt{{{3+\sqrt{5}}\over2}}
\biggr){{(\sqrt{5}+1)}\over{\sqrt{3+\sqrt{5}}}}
-ln\biggl(1+\sqrt{{{3-\sqrt{5}}\over2}}
\biggr){{(\sqrt{5}-1)}\over{\sqrt{3-\sqrt{5}}}}\biggr]\nonumber\\
& & ={1\over{8\pi M_\pi}}ln\biggl[{{1+\sqrt{5}}\over2}\biggr]
\end{eqnarray}
$I_{11}$ is a {\it new} integral that one needs to evaluate for
the DCX 1-loop graphs.
For values of  ${\cal P}^2\neq0,\ - M_\pi^2$, one gets the
following expressions, which are also  {\it new}:
\begin{eqnarray}
& & (i)\gamma_0(\omega=- M_\pi,\Omega=- M_\pi, {\cal P}^2= M_\pi^2-
\vec P^2)\nonumber\\
& & -\gamma_0(\omega= M_\pi,\Omega= M_\pi, 
{\cal P}^2= M_\pi^2-\vec P^2)\nonumber\\
& & =-{1\over{2\pi|\vec P|}}\Biggl({\rm arctan}\biggl[{M_\pi
\over{\sqrt{M_\pi^2+\vec P^2}}}\biggr]\nonumber\\
& & -{{|\vec P|}\over{\sqrt{M_\pi^2
+\vec P^2}}}
\sum_{k=0}^\infty {{(2k)!}\over{2^{2k}(k!)^2(2k+1)}}\nonumber\\
& & \times F_1\biggl({3\over2},k+1,1,2;
-{{\vec P^2}\over M_\pi^2},{M_\pi^2\over{[2M_\pi^2+
\vec P^2]}}\biggr)\Biggr),
\end{eqnarray}
where
\begin{eqnarray}
& & F_i(a,b,c,d;x,y)\equiv\rm Appell\ functions(i=1,2,3,4);\nonumber\\
& & F_1(a,b,c,d;x,y)=\sum_{m=0}^\infty\sum_{n=0}^\infty{{a_{m+n}b_mc_n}
\over{c_{m+n}}}{{x^my^n}\over{m!n!}};
\end{eqnarray}

\begin{eqnarray}
& & (ii) \gamma_0(\omega=- M_\pi,\Omega=0,{\cal P}^2=-\vec P^2)
={{\biggl[1+{{\vec P^2}
\over{2M_\pi^2}} -{{|\vec P|}\over M_\pi}
\sqrt{4+{{\vec P^2}\over M_\pi^2}}\biggr]}\over{16\pi^2
|\vec P|\sqrt{4+{{\vec P^2}
\over M_\pi^2}}}}\nonumber\\
& & \times\sum_{k=0}^\infty{{(2k)!}\over{(k!)^22^{2k}(2k+1)}}\nonumber\\
& & \times\biggl[2B(k+2,{1\over2})\ _2F_1\biggl(k+2,1;k+{5\over2};
{{\biggl[-{{\vec P^2}\over{2M_\pi^2}} 
+\sqrt{4+{{\vec P^2}\over M_\pi^2}}\biggr]}
\over{\biggl[-(1+{{\vec P^2}\over{2 M_\pi^2}})
+\sqrt{4+{{(\vec P^2}\over M_\pi^2}}\biggr]}}\biggr)\nonumber\\
& & -B(k+1,{1\over2})\ _2F_1\biggl(k+1,1;k+{3\over2};
{{\biggl[-{{\vec P^2}\over{2 M_\pi^2}} 
+\sqrt{4+{{\vec P^2}\over M_\pi^2}}\biggr]}
\over{\biggl[-(1+{{\vec P^2}\over{2 M_\pi^2}})
+\sqrt{4+{{\vec P^2}\over M_\pi^2}}\biggr]}}\biggr)\biggr]
\nonumber\\
& & -{1\over{16\pi^2|\vec P|\sqrt{4
+{{\vec P^2}\over M_\pi^2}}}}\sum_{k=0}^\infty
{{(2k)!}\over{(k!)^22^{2k}(2k+1)}}\nonumber\\
& & \times\int_0^1dx{{x^k(2x-1)}\over{\sqrt{1-x}}}
{1\over{\biggl(x-{{\biggl[(1+{{\vec P^2}\over{2 M_\pi^2}}
+\sqrt{4+{{\vec P^2}\over M_\pi^2}}
{{|\vec P|}\over{M_\pi}}\biggr]}
\over{{{\vec P^2}\over{2M_\pi^2}}
+\sqrt{4+{{\vec P^2}\over M_\pi^2}}
{{|\vec P|}\over{M_\pi}}\biggr]}}\biggr)}},
\end{eqnarray}

\vskip 0.5 true in

\subsection{Other Integrals}

The results obtained after evaluation of  
other integrals of $(B2) - (B3)$ are
given below.

(a) ${\bf 2\pi-Propagator\ Integrals}$

The following relations, (B20), (B22), (B24),
(B26) and (B28), are not explicitly given in the literature. These relations
are absolutely general 
and become relevant when evaluating the
off-threshold 1-loop amplitudes for kinematic points away from
(\ref{eq:kinptdef}).
\begin{equation} I_9\equiv{\cal J}^{\pi\pi}_2(0,M_\pi^2)
=-{1\over2}\Delta_\pi(M^2, 0);
\end{equation}
\begin{equation}
I_{10}\equiv
{\cal J}^{\pi\pi}_3(0,M_\pi^2)
={1\over3}{\cal J}^{\pi\pi}(0, M_\pi^2)\equiv{1\over 3}I_8;
\end{equation}

\begin{eqnarray} & & {\cal J}^{\pi\pi}_1({\cal P}^2, M^2)
= {1\over{2{\cal P}^2}}\biggl[
\Delta_\pi(0, M^2)-\Delta_\pi(0, M_\pi^2)
+(\Delta+{\cal P}^2){\cal J}({\cal P}^2, M^2)\biggr];\nonumber\\
& & I_2\equiv {\cal J}^{\pi\pi}_1(-M_\pi^2,M_\pi^2)
={1\over 2}{\cal J}^{\pi\pi}(-M_\pi^2,M_\pi^2)
\nonumber\\
& & = -L -{1\over{32\pi^2}}\biggl[-1+ln{M_\pi^2\over\mu^2}
+\sqrt{5} ln\biggl({{\sqrt{5}+1}\over{\sqrt{5}-1}}\biggr)\biggr]
\end{eqnarray}
\begin{eqnarray}
& & -M_\pi^2 I_5\equiv 
-M_\pi^2{\partial\over{\partial M^2}}{\cal J}^{\pi\pi}_1
|_{M^2=M_\pi^2;{\cal P}^2=-M_\pi^2} 
\nonumber\\
& & ={1\over 2}\biggl({\partial\over{\partial M^2}}
\Delta_\pi(0,M^2)-M_\pi^2{\partial\over{\partial M^2}}
{\cal J}^{\pi\pi}\biggr)_{M^2=M_\pi^2;{\cal P}^2=-M_\pi^2}
\nonumber\\
& & + {1\over 2}{\cal J}^{\pi\pi}(-M_\pi^2,M_\pi^2)
\\
& & =-{L\over  2}+{3\over{32\pi^2}}-{\sqrt{5}\over{40\pi^2}}
ln\biggl({{\sqrt{5}+1}\over{\sqrt{5}-1}}\biggr)
\end{eqnarray}

\begin{eqnarray} & & {\cal J}^{\pi\pi}_2({\cal P}^2; M^2= M_\pi^2)
=
{{\biggl[\biggl(2 M_\pi^2-{{\cal P}^2\over2}\biggr)J^{\pi\pi}({\cal P}^2)
-\Delta_\pi(0, M_\pi^2)
\biggr]}\over{2(d-1)}},\\
& & {\rm which\ for}\ {\cal P}^2=-M_\pi^2\ {\rm gives}:\nonumber\\
& & I_3\equiv {\cal J}^{\pi\pi}_2(-M_\pi^2,M_\pi^2)
\nonumber\\
& & = -{{7M_\pi^2}\over6}L+{{29}\over{576\pi^2}}M_\pi^2-
{7\over{192\pi^2}}M_\pi^2ln{M_\pi^2\over\mu^2}
-{{5\sqrt{5}}\over{192\pi^2}}
ln\biggl({{\sqrt{5}+1}\over{\sqrt{5}-1}}\biggr)+{\cal O}(d-4);
\nonumber\\
\end{eqnarray}
\begin{eqnarray}
& & {\cal J}^{\pi\pi}_3({\cal P}^2; M^2= M_\pi^2)\equiv
J^{\pi\pi}_3(P^2)\nonumber\\
& & ={{\biggl[\biggl(d {\cal P}^2-4 M_\pi^2\biggr)
J^{\pi\pi}({\cal P}^2)
+2(2-d)\Delta_\pi(0, M_\pi^2)\biggr]}
\over{4(d-1){\cal P}^2}},\\
& & {\rm which\ for}\ {\cal P}^2=-M_\pi^2\ {\rm gives}:
\nonumber\\
& & I_4\equiv{\cal J}^{\pi\pi}_3(-M_\pi^2,M_\pi^2)
\nonumber\\
& & = -{2\over3}L+{{19}\over{288\pi^2}}
-{1\over{48\pi^2}}ln{M_\pi^2\over\mu^2}-
{{\sqrt{5}}\over{24\pi^2}}ln\biggl({{\sqrt{5}+1}\over{\sqrt{5}-1}}\biggr)\nonumber\\
& & +{\cal O}(d-4).
\end{eqnarray}

\begin{eqnarray}
& & \theta(-{\cal P}^2){\cal P}^2
{\partial\over{\partial M^2}}{\cal J}^{\pi\pi}_2
({\cal P}^2; M^2) \nonumber\\
& & =\theta(-{\cal P}^2){1\over{2(1-d)}}
\Biggl[-{\cal J}^{\pi\pi}-{1\over 2}(4M_\pi^2-{\cal P}^2)\
{\partial\over{\partial M^2}}{\cal J}^{\pi\pi}+{1\over2}
{\partial\over{\partial M^2}}\Delta_{\pi}(0, M^2)\Biggr]
\\
&  & =-{L\over2}-{1\over{192\pi^2}}(1-{{\cal P}^2\over M_\pi^2})
-{1\over{64\pi^2}}ln{M_\pi^2\over\mu^2}\nonumber\\
& & -{1\over{64\pi^2}}\sqrt{{{[4M_\pi^2-{\cal P}^2]}\over{-{\cal P}^2}}}
ln\biggl[{{\sqrt{4M_\pi^2-{\cal P}^2}
+\sqrt{-{\cal P}^2}}\over{\sqrt{4M_\pi^2-{\cal P}^2}-
\sqrt{-{\cal P}^2}}}\biggr]+{\cal O}(d-4);\nonumber\\
& & {\rm So},\nonumber\\
& & I_6\equiv{\partial\over{\partial M^2}}
{\cal J}^{\pi\pi}_2(-M_\pi^2,M_\pi^2)|_{M^2=M_\pi^2}
= -{L\over 2}-{1\over{96\pi^2}}-{1\over{64\pi^2}}ln{M_\pi^2\over\mu^2}
-{\sqrt{5}\over{64\pi^2}}ln
\biggl({{\sqrt{5}+1}\over{\sqrt{5}-1}}\biggr);\nonumber\\
& & \nonumber\\
\end{eqnarray}

\begin{eqnarray}
& & M_\pi^2{\partial\over{\partial M^2}}
{\cal J}^{\pi\pi}_3|_{M^2=M_\pi^2}\nonumber\\
& & =
\biggl(d{\partial\over{\partial M^2}}{\cal J}^{\pi\pi}_2 - M^2
 {\partial\over{\partial M^2}}{\cal J}^{\pi\pi}\biggr)|_{M^2
= M_\pi^2} - {\cal J}^{\pi\pi};\\
& & {\rm So},\nonumber\\
& & I_7\equiv{\partial\over{\partial M^2}}
{\cal J}^{\pi\pi}_3(-M_\pi^2, M_\pi^2)|_{M^2=M_\pi^2}
= -{7\over{96\pi^2}}+
{1\over{16\sqrt{5}}}ln\biggl({{\sqrt{5}+1}\over{\sqrt{5}-1}}\biggr)
.\nonumber\\
& & 
\end{eqnarray}

(b) ${\bf 2\pi-1Nucleon-Propagator\ Integrals}$

The following relations, (B30), (B32)
and (B35), are not explicitly given in the literature, and
are absolutely general
and become relevant when evaluating the
off-threshold 1-loop amplitudes for kinematic points away from
(\ref{eq:kinptdef}).
\begin{eqnarray}
& & \gamma_1(\omega,\Omega,{\cal P}^2)
={{\biggl[2{\cal P}^2J^{\pi\pi}({\cal P}^2)+{\cal P}^2
(2\omega-\Omega)\gamma_0(\omega,\Omega,{\cal P}^2)-\Omega\biggl(J^{\pi N}
(\omega)-J^{\pi N}(\omega-\Omega)\biggr)
\biggr]}\over{2({\cal P}^2-\Omega^2)}};
\nonumber\\
& & \gamma_2(\omega,\Omega,{\cal P}^2)
={{\biggl[J^{\pi N}(\omega)-J^{\pi N}(\omega-
\Omega)-2\Omega J^{\pi\pi}({\cal P}^2)
+({\cal P}^2-2\omega\Omega)\gamma_0(\omega,\Omega,{\cal P}^2)
\biggr]}\over{2({\cal P}^2-\Omega^2)}}
\end{eqnarray}
Hence,
\begin{equation} I_{12}\equiv
\gamma_2(0,-M_\pi,0)-\gamma_2(0,M_\pi,0)
= {1\over{4M_\pi\pi^2}}
\end{equation}
\begin{eqnarray}
& & \gamma_3(\omega,\Omega,{\cal P}^2)
={{\biggl[2M_\pi^2\gamma_0(\omega,\Omega,{\cal P}^2)
-2\omega\gamma_1(\omega,\Omega,{\cal P}^2)
-P^2\gamma_2(\omega,\Omega,{\cal P}^2)-
J^{\pi N}(\omega-\Omega)\biggr]}\over{2(d-2)}}\nonumber\\
& & ={1\over4}\biggl[2M_\pi^2\gamma_0(\omega,\Omega,{\cal P}^2)
-2\omega\gamma_1(\omega,\Omega,P^2)-P^2\gamma_2(\omega,\Omega,{\cal P}^2)-
J^{\pi N}(\omega-\Omega)\biggr]\nonumber\\
&  & -{{(2\omega-\Omega)}\over{32\pi^2}}+{\cal O}(d-4).
\end{eqnarray}
Hence,
\begin{eqnarray}
& &  \gamma_3(\omega=0,\Omega=-M_\pi,{\cal P}^2=M_\pi^2-
\vec P^2)-\gamma_3(\omega=0,\Omega= M_\pi,{\cal P}^2= M_\pi^2-
\vec P^2)
\nonumber\\
& & ={1\over4}\biggl[\biggl(\gamma_0(0,-M_\pi, M_\pi^2
-\vec P^2)-\gamma_0(0, M_\pi, M_\pi^2
-\vec P^2)\biggr)
\times{{
[M_\pi^2+\vec P^2]^2}
\over{2\vec P^2}}\nonumber\\
& & +{{[M_\pi^2+\vec P^2]}\over{2\vec P^2}}
\biggl(J^{\pi N}(-M_\pi)-J^{\pi N}(M_\pi)\biggr)\biggr]
\nonumber\\
& & -{M_\pi\over{16\pi^2}}+{\cal O}(d-4).
\end{eqnarray}
Thus:
\begin{eqnarray}
& & I_{13}\equiv
\gamma_3(0, -M_\pi, 0)-\gamma_3(0, M_\pi,0)\nonumber\\
& & = 2L M_\pi+ M_\pi\biggl({1\over{64}}-{1\over{8\pi^2}}\biggr)
+{M_\pi\over{16\pi^2}}ln{M_\pi^2\over{\mu^2}}+{\cal O}(d-4).
\end{eqnarray}
Now,
\begin{eqnarray}
& & \gamma_5(\omega,\Omega,{\cal P}^2)
=-{1\over{2(d-2)({\cal P}^2-\Omega^2)}}
\biggl[(d-2)\Omega J^{\pi\pi}({\cal P}^2)\nonumber\\
& & +2\gamma_2(\omega,\Omega,{\cal P}^2)
[(d-1){\cal P}^2+\omega\Omega(d-2)]\nonumber\\
& & +(d-3)J^{\pi N}(\omega-\Omega)-2\omega
\gamma_1(\omega,\Omega,{\cal P}^2)
+2M_\pi^2\gamma_0(\omega,\Omega,{\cal P}^2)\biggr]
\nonumber\\
& & ={1\over{4({\cal P}^2-\Omega^2)}}\biggl[2\omega\gamma_1(\omega,\Omega,
{\cal P}^2)-2\Omega J^{\pi\pi}
+(3{\cal P}^2-4\omega\Omega)\gamma_2(\omega,\Omega,{\cal P}^2)
-J^{\pi N}(\omega-\Omega)\nonumber\\
& & -2M_\pi^2\gamma_0(\omega,\Omega,{\cal P}^2)
+{{(2\omega-\Omega)}\over{8\pi^2}}\biggr] + {\cal O}(d-4).
\end{eqnarray}

Hence,
\begin{eqnarray}
& & \gamma_5(\omega=0,\Omega=-M_\pi,{\cal P}^2= M_\pi^2-
\vec P^2)
-\gamma_5(\omega=0,\Omega= M_\pi,{\cal P}^2= M_\pi^2-
\vec P^2)\nonumber\\
& & 
= -{1\over{4\vec P^2}}\biggl[-{{[3M_\pi^2-5\vec P^2]}\over{2\vec P^2}}
\biggl(J^{\pi N}
(-M_\pi)-J^{\pi N}(M_\pi)\biggr)\nonumber\\
& & -{{(3[M_\pi^2-\vec P^2]^2+4M_\pi^2
\vec P^2)}\over{2\vec P^2}}
\biggl(\gamma_0(0,-M_\pi,M_\pi^2-\vec P^2)
-\gamma_0(0, M_\pi, M_\pi^2-\vec P^2)\biggr)
\nonumber\\
& & +{{2 M_\pi[5 M_\pi^2 - 3\vec P^2]}
\over{\vec P^2}}{\cal J}^{\pi\pi}(M_\pi^2-\vec P^2)
+{M_\pi\over{4\pi^2}}\biggr]+{\cal O}(d-4).
\end{eqnarray}
Thus:
\begin{eqnarray} & & I_{14}\equiv 
\gamma_5(0, -M_\pi, 0)-\gamma_5(0,M_\pi, 0)
\nonumber\\
& & ={1\over{16\pi^2M_\pi}}+{1\over{64M_\pi}}+{\cal O}(d-4).
\end{eqnarray}

\vfill\eject

\begin{figure}[p]
\centerline{\mbox{\psfig{file=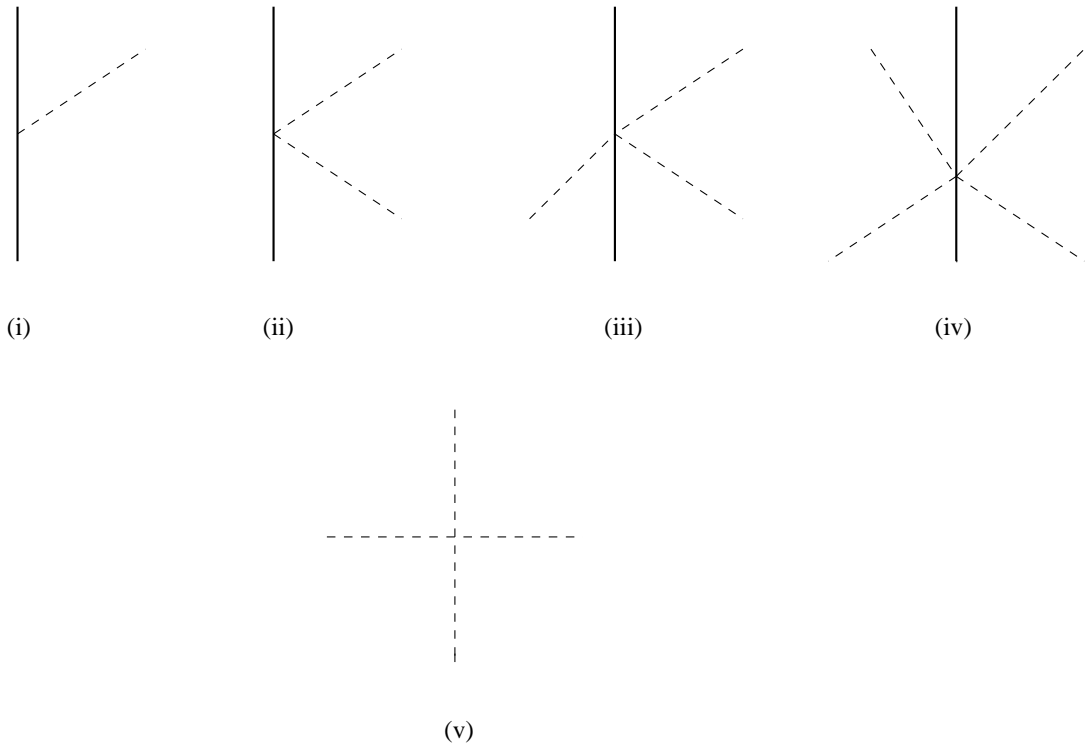,width=0.8\textwidth}}}
\caption{Elementary Vertices}
\end{figure}

\begin{figure}[p]
\centerline{\mbox{\psfig{file=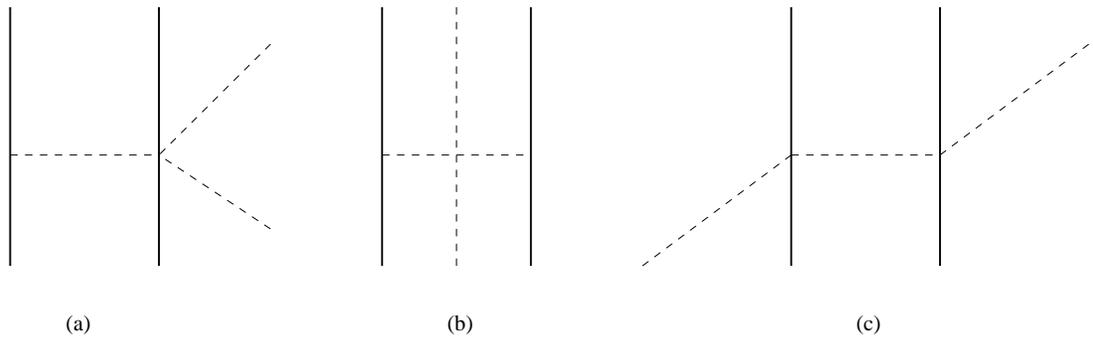,width=0.8\textwidth}}}
\caption{Tree Graphs}
\end{figure}

\begin{figure}[p]
\centerline{\mbox{\psfig{file=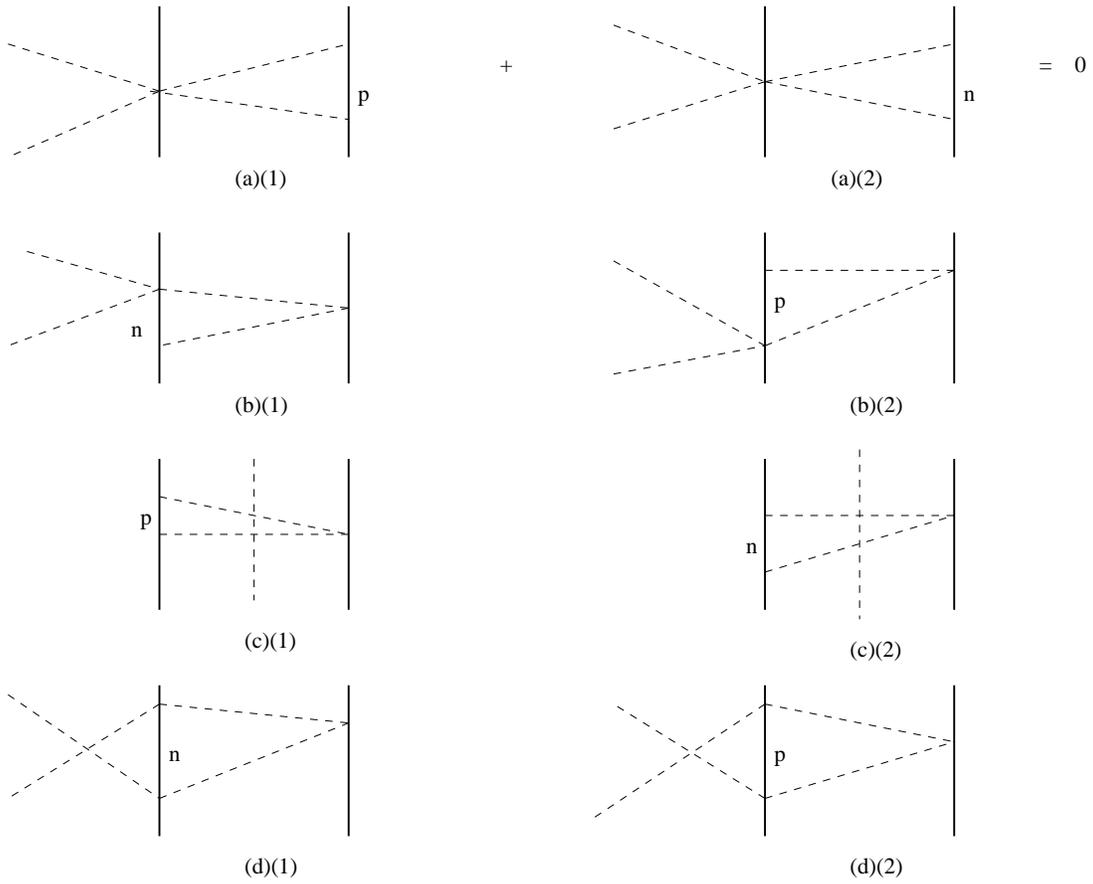,width=0.8\textwidth}}}
\caption{2Nucleon-1 Loop Graphs (a) -(d)}
\end{figure}

\begin{figure}[p]
\centerline{\mbox{\psfig{file=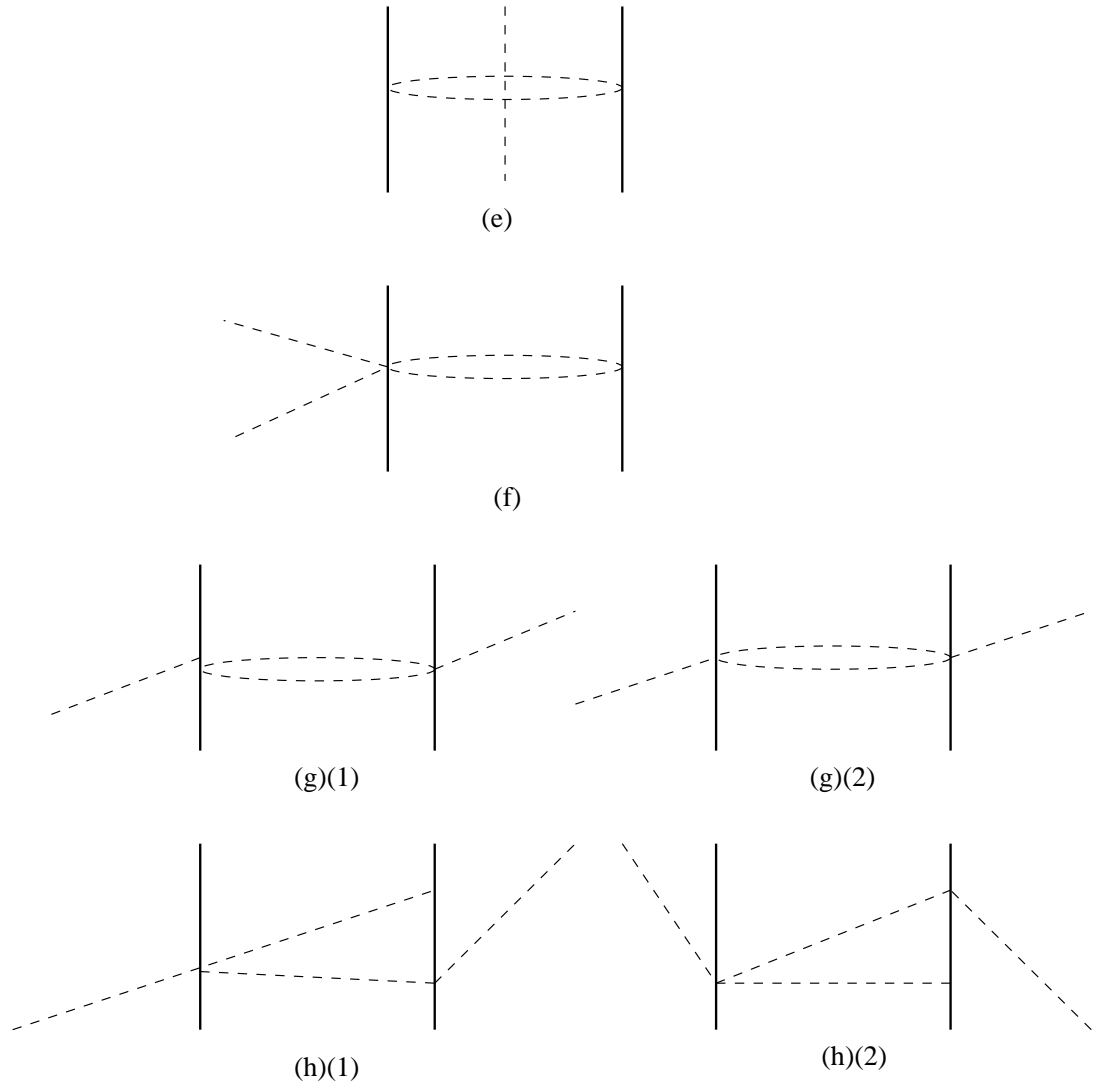,width=0.8\textwidth}}}
\caption{2Nucleon-1 Loop Graphs (e) - (h)}
\end{figure}

\begin{figure}[p]
\centerline{\mbox{\psfig{file=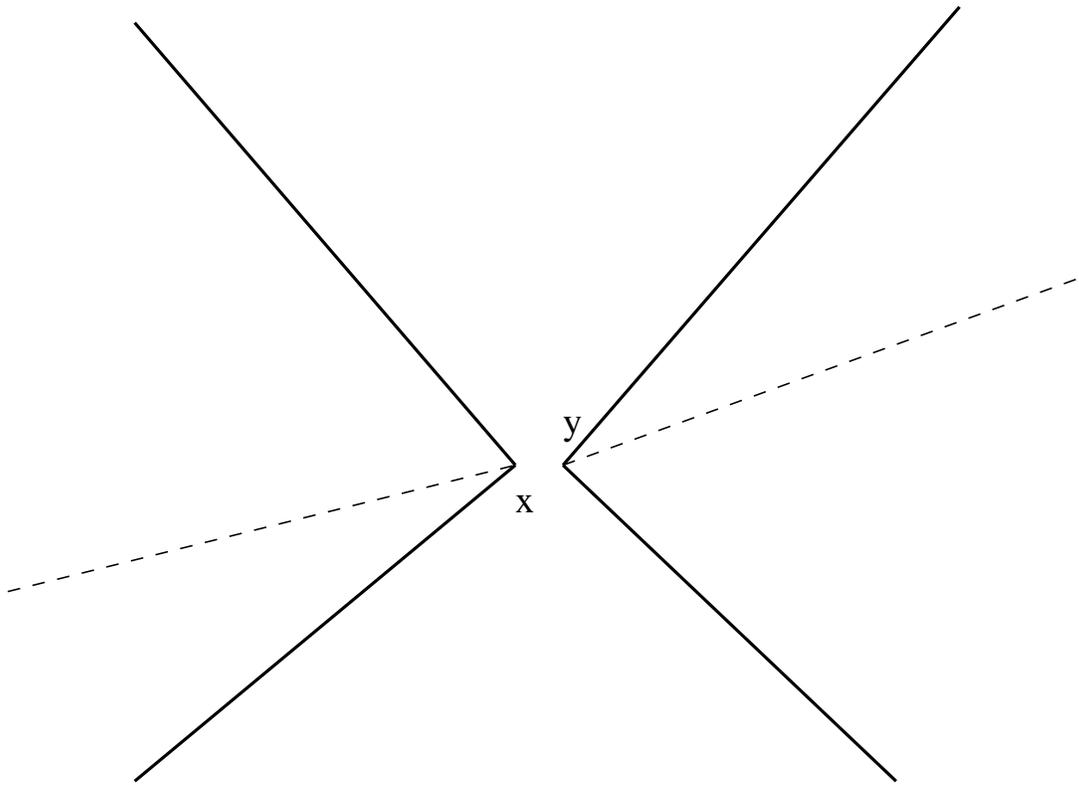,width=0.8\textwidth}}}
\caption{2$\pi$-2 Nucleon Contact Graph}
\end{figure}

\end{document}